\documentclass[fleqn,usenatbib]{mnras}

\usepackage[T1]{fontenc}
\usepackage{ae,aecompl}

\usepackage{graphicx}	
\usepackage{amsmath}	
\usepackage{amssymb}	

\newcommand{\ramses}{{\small RAMSES} }
\newcommand{\HI}{{\rm H}${\rm\scriptstyle I}$ }
\newcommand{\HIalt}{{\rm H}${\rm\scriptstyle I}$}
\newcommand{\Halpha}{{\rm H$\alpha$} }
\newcommand{\Halphaalt}{{\rm H$\alpha$}}
\newcommand{\Msol}{\,{\rm M}_\odot}
\newcommand{\Msolyr}{{\rm \,M}_\odot\,{\rm yr}^{-1}}
\newcommand{\kmsec}{{\,\rm {km\,s^{-1}} }}

\newcommand{\Qgtext}{$Q_{\rm g}$ }

\newcommand{\sigmag}{$\sigma_{\rm g}$\ }
\newcommand{\sigmagalt}{$\sigma_{\rm g}$}
\newcommand{\sigmasfr}{SFR-$\sigma_{\rm g}$ }

\title[Star formation and turbulence in the ISM]{From giant clumps to clouds - III. The connection between star formation and turbulence in the ISM} 
\author[Timmy Ejdetj{\"a}rn et al.]{
	Timmy Ejdetj{\"a}rn$^{1,2}$,
	Oscar Agertz$^{2}$,
	G\"oran \"Ostlin$^{1}$,
	Florent Renaud$^{2}$
	and Alessandro B. Romeo$^{3}$
	\\
	$^{1}$Oskar Klein Centre, Department of Astronomy, Stockholm University, 106 91 Stockholm, Sweden\\
	$^{2}$Department of Astronomy and Theoretical Physics, Lund Observatory, Box 43, SE-221 00 Lund, Sweden\\
	$^{3}$Department of Space, Earth and Environment, Chalmers University of Technology, SE-41296 Gothenburg, Sweden
}

\date{Accepted XXX. Received YYY; in original form ZZZ}

\pubyear{2022}

\usepackage{newtxtext,newtxmath}

\begin{document}
	\label{firstpage}
	\pagerange{\pageref{firstpage}--\pageref{lastpage}}
	\maketitle
	
	\begin{abstract}
		Supersonic gas turbulence is a ubiquitous property of the interstellar medium. The level of turbulence, quantified by the gas velocity dispersion (\sigmagalt), is observed to increase with the star formation rate (SFR) of a galaxy, but it is yet not established whether this trend is driven by stellar feedback or gravitational instabilities. In this work we carry out hydrodynamical simulations of entire disc galaxies, with different gas fractions, to understand the origins of the \sigmasfr relation. We show that disc galaxies reach the same levels of turbulence regardless of the presence of stellar feedback processes, and argue that this is an outcome of the way disc galaxies regulate their gravitational stability. The simulations match the \sigmasfr relation up to SFRs of the order of tens of $\Msolyr$ and $\sigma_{\rm g}\sim 50\kmsec$ in neutral hydrogen and molecular gas, but fail to reach the very large values ($>100\kmsec$) reported in the literature for rapidly star-forming galaxies. We demonstrate that such high values of $\sigma_{\rm g}$ can be explained by 1) insufficient beam smearing corrections in observations, and 2) stellar feedback being coupled to the ionised gas phase traced by recombination lines. Given that the observed \sigmasfr relation is composed of highly heterogeneous data, with \sigmag at high SFRs almost exclusively being derived from \Halpha observations of high redshift galaxies with complex morphologies, we caution against analytical models that attempt to explain the \sigmasfr relation without accounting for these effects.
	\end{abstract}

	
	

	\section{Introduction}
    The interstellar medium (ISM) of disc galaxies is observed to be highly dynamic and complex in nature. A striking property of the ISM is its supersonic turbulence across a wide range of spatial scales, redshifts, and a multitude of gas tracers \citep[see e.g.][for reviews]{ElmegreenScalo04, MacLowKlessen04, Glazebrook13}. The presence of supersonic turbulence has been found to have a crucial impact on the temperature and density distribution in the galaxy \citep[e.g.][]{McKeeOstriker07}, gas mixing \citep[][]{YangKrumholz12, ArmillottaKrumholz18}, and the formation of stars \citep[][]{MacLowKlessen04, Ballesteros-Paredes+07, Renaud+12, Padoan+14, Federrath2018}.
    
    The dynamics of the ISM in galaxies through a large range of evolutionary stages is well observed with several tracers, corresponding to different distinct gas phases. The level of turbulent gas motions, commonly quantified as the velocity dispersion \sigmag (measured as the line width of individual emission lines), can vary significantly between phases. In the warm ionised phase of local disc galaxies, it is of the order of $\sigma_{\rm g}\sim 10-40\kmsec$ \citep[e.g.][]{Moiseev+15, Varidel+16} and it rises to $\gtrsim 100\kmsec$ in rapidly star forming high-redshift galaxies \citep[e.g.][]{Epinat+09, Cresci+09, Law+09, Genzel+11, Alcorn+18}. In contrast, in local disc galaxies, the velocity dispersion in atomic hydrogen \citep{Ianjamasimanana+12, Stilp+13} and molecular hydrogen \citep[e.g.][]{Caldu-Primo+13, Nguyen-Luong+16, Levy+18, Girard+21} is commonly observed around $\sigma_{\rm g}\sim 10\kmsec$. 
    
    A relation between the star formation rates (SFRs) of galaxies and \sigmag is observed across several observed scales, redshifts and gas tracers \citep[e.g.][]{Epinat+09, Ianjamasimanana+12, Moiseev+15, Alcorn+18, Levy+18}. This implies that the process of star formation is accompanied by the injection of turbulence in the ISM. However, there is no consensus on what drives and maintains the ISM turbulence. A large number of candidates have been argued in literature \citep[see ][for a review]{ElmegreenScalo04, Glazebrook13}, with some combination of stellar feedback and gravitational instabilities likely being important \citep[e.g.][]{Dib06,Agertz+09b, Lehnert+09, KBFC18, Orr+19}. Stellar feedback intuitively fits the scenario, as supernovae and stellar winds naturally insert large amounts of energy and momentum into the ISM, and has a close connection to star formation. However, gravitational instabilities are also closely connected to the formation of stars and can convert potential energy into turbulent energy, e.g. in the form of clump formation, accretion or radial flows through the disc \citep[e.g.][]{Dekel+09a, Agertz+09b, KrumholzBurkert10}. Cosmological processes, such as accretion and mergers, have also been suggested to directly induce turbulence during interactions \citep[e.g.][]{Forster-Schreiber+06, Dekel+09b, KlessenHennebelle2010, Genel+12a, Renaud+14}. However, accretion might only be relevant early-on in disc galaxies before feedback and instabilities become dominant \citep[see][and references therein]{ElmegreenBurkert2010}. \citet{Ginzburg+22} found that accretion is unlikely the primary driver of turbulence in less massive halos $<10^{12}\,\Msol$ (at $z=0$).

	\begin{figure*}
		\centering
		\includegraphics[width=1.0\textwidth]{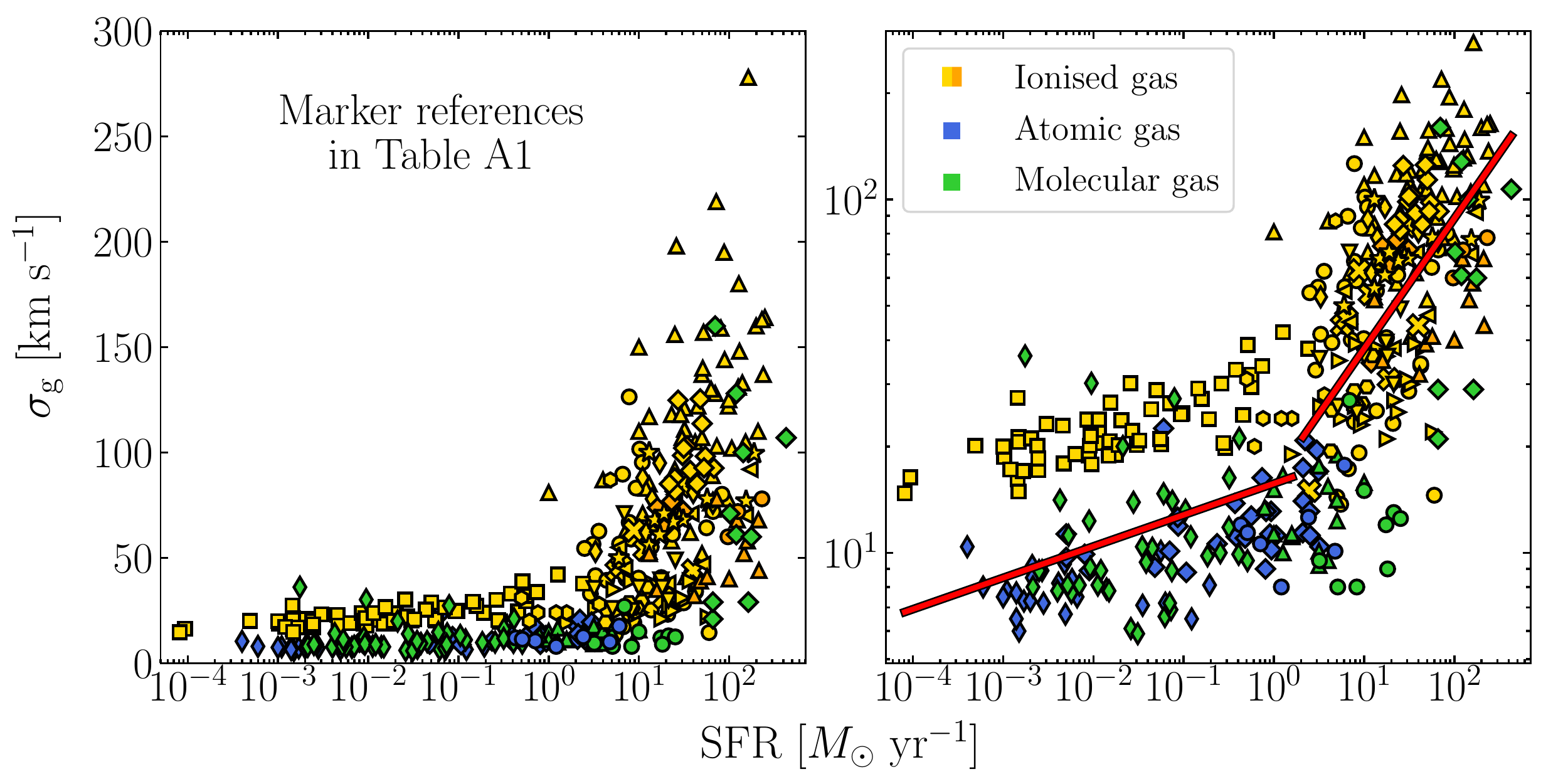}
		\caption{ Observational data of the relation between the gas velocity dispersion $\sigma_{\rm g}$ (a measurement on the level of gas turbulence) and the SFR. The markers represent individual galaxies (sources and additional information shown in Table~\ref{tab:data}) and are colour-coded by the gas phase of the tracer used to measure \sigmagalt. Left figure shows the relation as it is commonly evaluated in literature and the right figure shows the relation in log-space in order to highlight the kinematic difference at some threshold SFR$\,\sim 2\Msolyr$; further illustrated by the two fitted (red) lines. The data set is highly heterogeneous and the details of this are discussed in Section~\ref{sec:observations}.}
		\label{fig:data}
	\end{figure*}
	
	Several authors have developed analytic models of disc galaxies which incorporate \sigmag and can thus be used to evaluate the source of galactic turbulence \citep[e.g.][]{FQH13,KrumholzBurkhart16, HaywardHopkins17, KBFC18, Orr+19, NusserSilk21}, which usually focus on modelling one particular possible origin. \citet{KrumholzBurkhart16} presented two galaxy-wide analytic models of galactic discs designed to reproduce the \sigmasfr relation. One of their models is the analytic steady-state solution of a galactic disc derived by \citet{KrumholzBurkert10}, which is based on gravitational instabilities driving turbulence through radial inflows through the disc. They compare this model with a feedback-driven model by \citet{FQH13}, which assumes that the vertical gas pressure is balanced by the momentum input from supernovae explosions. \citet{KBFC18} extended this study by deriving a model that incorporates both effects, meaning that it can be applied to a wide diversity of galaxies. In recent years, several observational studies have applied these models to interpret observational data \citep[][]{Yu+19, Ubler+19, Varidel+20, Girard+21, Yu+21}. While such analytic models are efficient and useful for comparisons with data, they tend to involve a large set of free parameters that are not straight forward to constrain observationally. 
	
	
    Recent cosmological simulations have managed to reproduce the \sigmasfr relation in disc galaxies over cosmic time (\citealt[][]{Hung+19, Orr+20}; or the $\Sigma_{\rm SFR}$-$\sigma_{\rm g}$ relation in the case of Orr et al. 2020). \citet[][]{Orr+20} showed that the most tenuous gas was also the most turbulent, which highlights the kinematic differences between gas phases. They also compared their simulations to analytic models and concluded that both a feedback-regulated model and marginal (gravitational) stability can explain the turbulence in the neutral gas of their galaxies. However, these studies did not evaluate in detail the impact of observational artefacts on the \sigmasfr relation.
	
	In this paper, we present hydrodynamical simulations of entire disc galaxies of varying gas fractions and SFRs. We use these simulations to understand the roles played by stellar feedback and disc gravitational instability, as well as observational biases, in shaping the observed \sigmasfr relation. The paper is organised as follows. In Section~\ref{sec:observations} we present the compilation of observational data considered in this work. In Section~\ref{sec:simulations} we present the numerical code used for our simulations and how star formation and stellar feedback processes are implemented. In Section~\ref{sec:results} we present the galaxy simulations and the emerging \sigmasfr relation in each simulated galaxy. Having done so, we quantify the impact of beam smearing and the considered gas phases/tracers on the observed velocity dispersion. In Section~\ref{sec:discussion} we evaluate modern analytic models of galactic disc turbulence compare to the hydrodynamical simulations, and discuss the caveats that exists when confronting them to observations. Finally, we carry out a stability analysis of the galaxies and demonstrate the ability of the $Q$ stability parameter to predict the simulated levels of turbulence. Our conclusions are summarised in Section~\ref{sec:conclusions}.

	\section{Observational data and biases}
	\label{sec:observations}
	Figure~\ref{fig:data} presents the observational data for the star formation rate (SFR) and gas velocity dispersion (\sigmagalt) that we consider in this work. Each marker represents an individual galaxy\footnote{With the exception of data taken from \citet{Nguyen-Luong+16} who observed individual star forming molecular clouds within the Milky Way.} and is colour-coded according to the phase of the gas tracer of \sigmagalt; see Table~\ref{tab:data} for references and general information about the observations. The left panel shows the relation in semi-log space \citep[as is commonly done in the literature, e.g.][]{KrumholzBurkhart16,KBFC18}, and the right panel in log-space. The \sigmasfr relation is commonly considered to be well-described by a single power law, but as can be seen from the fitted (red) lines in the right panel the relation better follows a piecewise power law which at low SFR goes as $\sigma_{\rm g} \propto {\rm SFR}^{0.1}$ and steepens to $\sigma_{\rm g} \propto {\rm SFR}^{0.4}$ at ${\rm SFR}\gtrsim2\Msolyr$. These fits show only an average slope of all the data and the relation may have further dependencies, e.g. gas phase or scale.
	
    There exists a number of important observational caveats related to this relation that need to be considered to form a more homogeneous data set that can be investigated into further detail. A disc galaxy observed at a high inclination will have its rotational motion almost directly along the line-of-sight (LoS), which widens the distribution of the observed radial velocities; thus increasing the velocity dispersion. For observations at low spatial resolution, spectral features at different LoS velocities become blended. This is the cause of a known observational effect called \emph{beam smearing}, which artificially increases the observed velocity dispersion, and is exacerbated for highly inclined galaxies where rotation blends together with turbulent motions.
    
    Several codes have been developed to correct for beam smearing by modelling a typical galaxy and reconstructing its rotation curve and kinematics \citep[e.g. ROTCUR, GALFIT, TiRiFiC, DYSMAL, 3D-Barolo;][respectively]{vanAlbada+85, Peng+02, Josza+07, Davies+11, DiTeodoroFraternali15}. The methods used vary between the codes, which can significantly affect the derived velocity dispersion. In particular, methods based on 2D modelling are biased towards higher \sigmag values \citep[see the discussion in][]{DiTeodoroFraternali15}, which \citet{Rizzo+21} highlighted as an under-prediction of $V/\sigma$. Assuming a constant velocity dispersion throughout the disc has also been shown to lead to a bias towards higher \sigmag estimate \citep[see the discussion in][]{Rizzo+21}. Instead, 3D models (making use of datacubes) need to be employed to accurately correct for beam smearing. Furthermore, disc modelling of near edge-on or thick discs, where the line-of-sight overlaps the disc several times, results in ambiguous velocity dispersions. At low inclinations ($<40^\circ$), tilted-ring models might have significant residual error due to the difficulty of disentangling the rotational velocity and the velocity dispersion \citep{Kamphuis+15}.


	\begin{table*}
		\centering  
		\caption{Characteristics of each simulation. Refinement time refers to the time allowed for the initial conditions to relax, after which the refinement/resolution is increased.}
		\begin{tabular}{l l l l l l l}
			\hline\hline
			Name of run &  $f_{\rm g}$ (\%) & Feedback? & $\epsilon_{\rm ff}$ (\%) & Refinement time (Myr) & SFR$^1$ ($\Msolyr$) \\
			\hline 
			\texttt{fg10\_noFB} & 10  & No & 1 & 200 & 4 \\
			\texttt{fg10\_FB} &  10 & Yes & 10 & 200 & 1 \\
			\texttt{fg50\_noFB}  & 50 & No & 1 &  100 & 100 \\
			\texttt{fg50\_FB}  & 50 & Yes & 10 &  100 & 25 \\
			\hline
		\end{tabular}
		\label{tab:runs}
		\\
		\hspace{-80mm} $^1$ 50 Myr after refinement
	\end{table*}

	Observational data from high-redshift galaxies ($z\gtrsim 1$) suffer from poor spatial resolutions. This makes comparison between observational data of turbulence in high-redshift and local galaxies troublesome, due to possible scaling relations between turbulence and the observed scale. Galaxy observations exploring the scaling between \sigmag and spatial resolution report conflicting results and are not conclusive, but on average an increase in \sigmag is expected with an increase of scale \citep[see][and references therein]{ElmegreenScalo04}. We explore the role of observed scale in Section~\ref{sec:scale} and find that for a wide range of scale, if the galaxy is not inclined, \sigmag differs by a factor of, at most, 2. However, if the galaxy is highly inclined ($>60^\circ$), and if this inclination is unaccounted for, patch size could have a major impact; even sufficient to explain the entire \sigmasfr relation! Furthermore, lower resolution naturally means fewer data points, which complicates the fitting process used in beam smearing corrections. Due to the range of orbital velocities present over a large spatial scale, the velocity field can not be accurately retrieved at high redshifts \citep[][]{Epinat+09}. However, even at low redshifts beam smearing may have a substantial effect on \sigmag from galaxies with a high SFR \citep[][]{Varidel+16}. \citet{Kohandel+20} employed cosmological simulations and identified that if beam smearing is not accounted for in high-z galaxies, the velocity dispersion can be overestimated by as much as over a factor of two.

	In the literature, \sigmag tends to be reported as a weighted average of the velocity dispersion calculated in each element (i.e. pixels), and then used to represent the entire galaxy; along with a galaxy-wide SFR, which is simply the sum of SFR from each pixel. Several different weights have been used in this regard: flux \citep[e.g.][]{Lemoine-Busserolle+10, Wisnioski+11, Varidel+16}, luminosity \citep[e.g.][]{Davies+11, Moiseev+15} and errors from emission line fitting \citep{Epinat+09}. \citet{Lehnert+13} briefly evaluated the significance in weighing with the surface (\Halphaalt) brightness, signal-to-noise and flux, but found that the resulting \sigmag were consistently only $\sim 10\%$ higher than the non-weighted mean. Several authors have suggested that a flux-weighted average favours the inner, more luminous, regions of galaxies \citep[][]{Davies+11, DiTeodoro+16}. These regions are also highly turbulent, which may result in brightness- or SFR-weighting yielding a bias towards large \sigmagalt. Furthermore, these inner regions are also more prone to be beam smeared, due to the steeper rotational curve velocity near the galaxy centre \citep[e.g.][]{Josza+07}.
		
	As mentioned in the introduction, the ISM has distinct gas phases with different dynamics. Thus, data from different observational tracers can not be directly compared, as the velocity dispersion of each phase can not be connected in a straightforward manner through theory. For example, Figure~\ref{fig:data} shows a clear offset of the \Halpha observational data at low SFRs (the yellow squares) when compared to other tracers. Different gas tracers could then trace separate parts of the \sigmasfr relation. This is further discussed in Section~\ref{sec:mock}.
	
	There is no universally accepted method to measure the velocity dispersion in galaxies, and it is thus difficult to form a completely homogeneous data set. For example, most authors have measured \sigmag within individual pixels and then determined a weighted mean \citep[e.g.][]{Epinat+09, Lemoine-Busserolle+10, DiTeodoro+16, Girard+21}, while some \citep[e.g.][]{Lehnert+13} have determined \sigmag from the integrated spectra (summing up all spectra before fitting the line width). The first method to some extent corrects for rotation and reduce (but do not eliminate) the effect of beam smearing. Others still have employed (empirical) models combined with observational data to determine \sigmag \citep[e.g.][]{Cresci+09}. \citet{Epinat+08} reported the dispersion in radial velocity, rather than line-of-sight velocity, from their best fit disc models (essentially the kinematic residuals of a well-behaved disc). The difference in method deployed by authors might give rise to systematic differences in \sigmag \citep[e.g.][attain \sigmag$>150\kmsec$, significantly higher than found by \citealt{Cresci+09, Lemoine-Busserolle+10} with similar data]{Lehnert+13}. Furthermore, there is a wide range of morphologies and galactic properties in the set of data we compare with here, as it contains local, high-redshift and dwarf galaxies. However, understanding these differences is important to correctly utilise the observational data.

	\section{Numerical method}\label{sec:simulations}
	In this work we perform hydrodynamic+{\it N}-body simulations of entire galactic discs using the Adaptive Mesh Refinement (AMR) code \ramses \citep{Teyssier02}. The code solves the Euler equations for the fluid dynamics using the Godunov scheme assuming an ideal mono-atomic gas with an adiabatic index $\gamma = 5/3$. Stars and dark matter are represented by collisionless particles. The accelerations of the particles and gas are computed from the gravitational potential, via the Poisson equation, using the multi-grid method \citep{GuilletTeyssier11} at each refinement level. Metal-dependent gas cooling follows the tabulated cooling functions of \citet{SutherlandDopita93} for $T > 10^4$ K and \citet{RosenBregman95} for lower temperatures.

	\subsection{Star formation and stellar feedback}
    The adopted star formation and feedback physics is presented in \citet[][see also \citealt{AgertzRomeoGrisdale15}]{Agertz+13}, which we briefly summarise here. Star formation is treated as a Poisson process, sampled using $10^3\Msol$ star particles on a cell-by-cell basis according to the star formation law, 
	\begin{align}\label{eq:star_formation}
	\dot\rho_\star = \epsilon_{\rm ff}\frac{\rho}{t_{\rm ff}}\ {\rm for}\ \rho>\rho_\star,
	\end{align}
	where $\rho$ is the gas density, $\rho_\star=100~m_{\rm H}{\rm cm}^{-3}$ is the adopted density threshold, $t_{\rm ff}=\sqrt{3\pi/32G\rho}$ is the free-fall time of a spherically symmetric cloud and $\epsilon_{\rm ff}$ is the star formation efficiency per free-fall time. In the simulations including stellar feedback we adopt $\epsilon_{\rm ff}=10\%$, which has been shown to be able to reproduce the properties of ISM and giant molecular cloud populations, as well as observed $\epsilon_{\rm ff}$. In the absence of star formation regulation via feedback, we adopt a lower $\epsilon_{\rm ff}=1\%$ as it better captures the mean $\epsilon_{\rm ff}$ in giant molecular cloud populations \citep[see][]{Grisdale2017,Grisdale2018,Grisdale2019}. Each formed star particle is assumed to be a simple stellar population with a \citet{Chabrier03} initial mass function.
	
	Central to our study is the ability of stellar feedback to drive ISM turbulence. The feedback model treats the time-dependent injection of momentum, energy, mass and heavy metals\footnote{We track iron  (Fe)  and  oxygen  (O)  abundances  separately, see \citep[][]{Rhodin2019}.} from supernovae (SNe) type Ia and type II, stellar winds and radiation pressure \citep[see][for details]{Agertz+13}. SNe are treated as discrete events \citep{AgertzRomeoGrisdale15}. To robustly capture the effect from SNe, which is the dominant source of feedback in terms of momentum and energy input, we follow \citet[][see also \citealt{Martizzi+2015}]{KimOstriker15} and inject the terminal SN momentum to ambient cells when the cooling radius is resolved by less than 6 cells. 

	\subsection{Simulation suite}
	\label{sec:simsuite}
	The initial conditions are based on those of the isolated disc galaxy in the {\small AGORA} project \citep{Kim+2014, Kim+2016}, set up to approximate a Milky Way-like galaxy following the methods described in \citet{Hernquist1993} and \citet{Springel2000}. Briefly, we adopt a NFW dark matter halo \citep[][]{NavarroFrenkWhite96} with a concentration parameter $c=10$ and virial circular velocity $v_{\rm 200}=150\kmsec$. This translates into a halo virial mass $M_{\rm 200}=1.1\times 10^{12}\Msol$ within $R_{\rm 200}=205$\,kpc. The total baryonic disc mass is $M_{\rm disc}=4.5\times10^{10}\Msol$ with a gas fraction that we vary, as described below. The initial stellar and gaseous components follow exponential surface density profiles with scale lengths $r_{\rm d}=3.4$\,kpc and scale heights $h=0.1r_{\rm d}$. The bulge-to-disc mass ratio is $0.125$. The bulge mass-profile follows a Hernquist profile \citep{Hernquist1990} with scale-length $0.1r_{\rm d}$. The dark matter halo and stellar disc are represented by $10^6$ particles each and the bulge consists of $10^5$ particles.

	In this work we are interested in modelling the evolution of high redshift and local disc galaxies. The key parameter which evolves over time and sets several important characteristics (e.g. star formation) of a galaxy is its gas fraction. By varying the initial gas fraction, at a fixed total disc mass, in the initial conditions, we model a `low redshift' galaxy, with $f_{\rm g}=10\%$, and the `high redshift' counterpart, with $f_{\rm g}=50\%$ \citep[see][for an explicit approach]{vanDonkelaar+21}. These models are not necessarily analogues of the same galaxy, but they still capture the environment inside galaxies at different evolutionary stages.

	Both gas fractions are simulated with and without stellar feedback processes (see Table~\ref{tab:runs} for details). This allows us to constrain the degree to which feedback or galaxy self-gravity drive ISM turbulence \citep[see also][]{Agertz+09b}. The galaxies are simulated in isolation, i.e. without cosmological context, in order to understand how ISM turbulence is driven by internal effects rather than environmental effects such as galaxy interactions and gas accretion. These additional effects perturb the morphology and kinematics of the disc and, thus, the stability and evolution of the disc. In particular, \citet[][]{Hafen+22} showed that the formation of thin discs, which are inherently less turbulent, is unfavorable when accretion is dominated by cold filaments. 

	We begin by evolving the initial conditions at low spatial resolution ($\Delta x\approx 150$\,pc) for 100-200 Myr (see Table~\ref{tab:runs}) in order for them to relax and the galaxy to develop spiral structure. After this epoch of letting the galaxies relax, we allow for maximum refinement and activate stellar feedback. Our analysis is done on the data outputs from the start of this refinement period until 200 Myr after, with a time between outputs of 5-25 Myr. We allow for the adaptive mesh to refine to a maximum of 16 levels, corresponding to a spatial resolution of $\Delta x \gtrsim 9$\,pc.
	
	 \begin{figure*}
		\centering
		\includegraphics[width=0.65\textwidth]{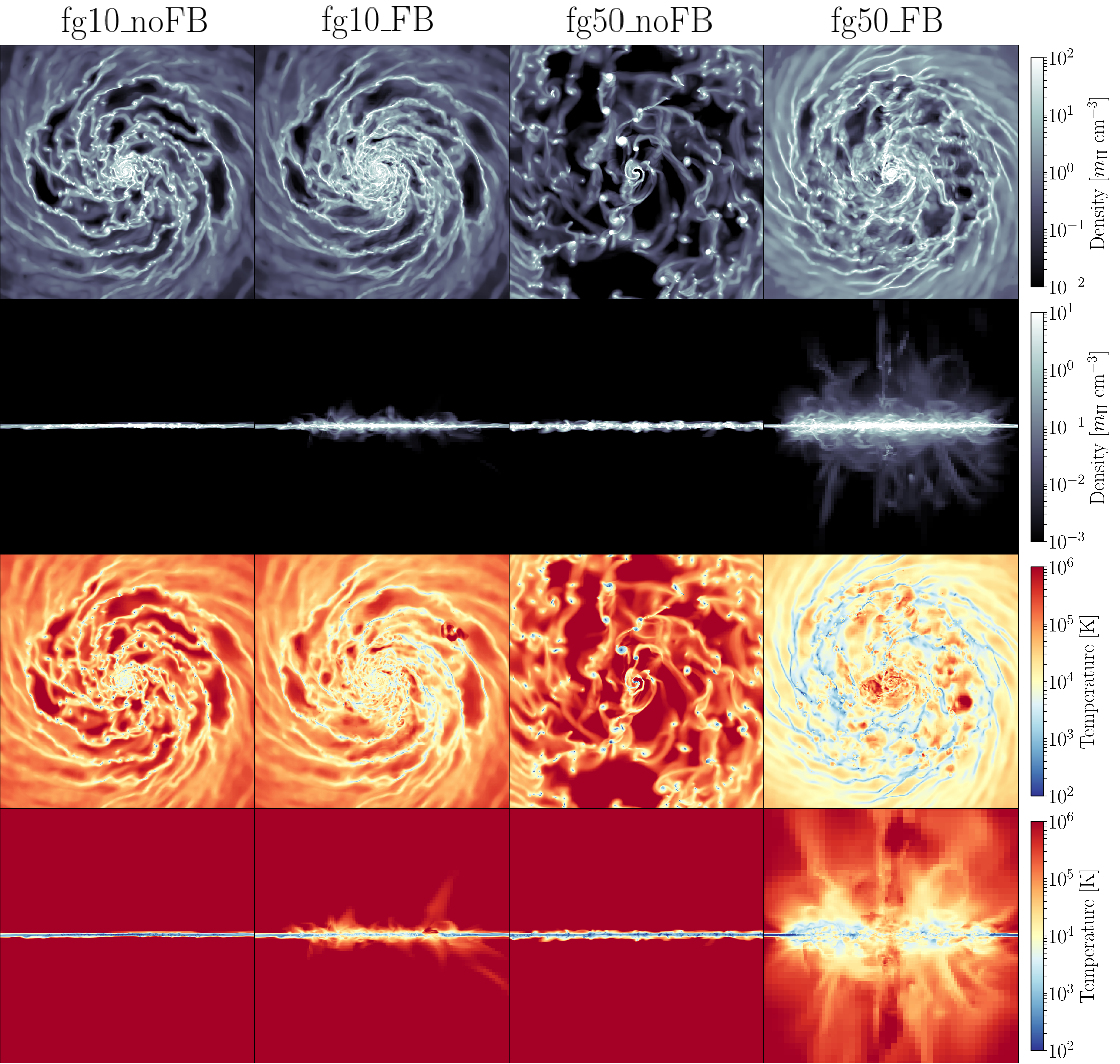}
		\caption{ The gas density and temperature in the simulated galaxies. The parameters were taken as the mass-weighted average of the cells along line of sight (face- and edge-on). The size of the boxes are 20 kpc\,$\times$\,20 kpc, and calculated 50\,Myr after feedback was turned on. The runs without any feedback are seen to form dense clumps of gas and there is a clear correlation with the denser areas being colder. Some clumping, as seen in the \texttt{fg50\_FB} run, is expected in high-redshift galaxies. Violent gas outflow caused by feedback are observed from the side-views and is more prominent in the more gaseous disc.} 
		\label{fig:visual}
	\end{figure*}
	
		    	    \begin{figure*}
		\centering
		\includegraphics[width=0.65\textwidth]{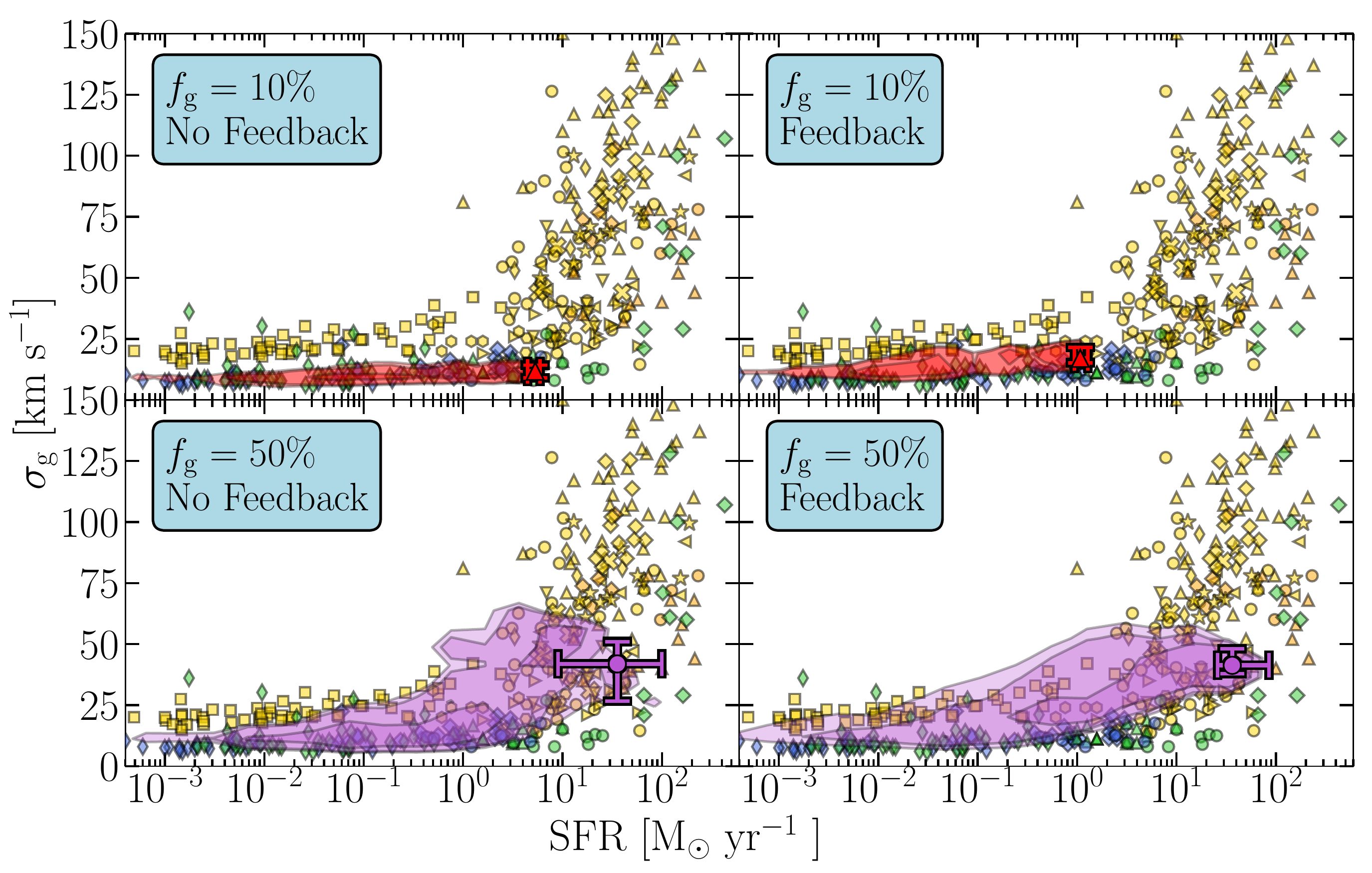}
		\caption{ The \sigmasfr relation for the galaxy simulation. The markers correspond to the literature data in Table \ref{tab:data}. The simulation data using a patch-approach is presented as a coloured contour and the global averages are shown as markers with error bars, as described in Section~\ref{sec:reduce_data}. The simulation data shown here is for no inclination, uses patch sizes $\geq 1$\,kpc, and includes all of the gas (regardless of phase). A remarkable agreement between simulations and observations can be seen and the simulations reproduce two distinct properties of the \sigmasfr relation: the plateau at SFR\,$\lesssim 2\,\Msolyr$ and (in the high gas fraction galaxy) the steep increase at higher SFRs. Feedback does not have a significant impact on the overall \sigmagalt, but we notice a higher dispersion in the values of \sigmag for all SFRs; this is related to the formation of the warm ionised phase (discussed in detail in Section~\ref{sec:mock}). }
		\label{fig:sigmasfr}
	\end{figure*}

    \subsection{Measuring the velocity dispersion and star formation rate}\label{sec:reduce_data}
    In order to simplify the comparison to literature data where galaxy-wide values for \sigmag and the SFR are frequently reported, we calculate these quantities within local patches and then combine these into global (galaxy-wide) averages. Patches allow us to evaluate how ISM properties depend on the observational resolution of specific observations, and hence physical scale - an important concept in astrophysical turbulence \citep[e.g.][]{MacLowKlessen04,kritsuk07,Grisdale2017}. Global quantities are particularly relevant in high-$z$ surveys where the spatial resolution becomes comparable to the size of the galaxy (as discussed in Section~\ref{sec:observations}).
    
    The local patches are analogue to a single resolution element of an observation, i.e. a pixel or spaxel, and are here defined as a set of cylindrical beams with diameters ranging between $0.1-10$\,kpc. The patches are distributed uniformly in the radial direction and with the same angular separation. Thus, the number of patches are concentrated near the galactic centre, the region of interest when investigating high SFRs and \sigmagalt. When investigating the effects of galaxy inclination, the beams were inclined relative to the disc plane according to our choice of observed line-of-sight angles: ($0,\,15,\,30,\,45,\,60,\,75,\,90^\circ$). The local setup then produces $\sim1000$ patches for each output (before rejecting unusable patches, i.e. with no gas mass or no star formation).

    Each patch covers several simulation cells. Every local quantity was calculated by summing or averaging the values of each simulation cell within a patch. Thus, the local SFR was calculated as the sum of SFR of all cells within a patch. The SFR within a patch was in turn calculated by binning the stellar ages of stars in bin sizes of 10 Myr (motivated by the lifetime of HII regions), then calculating the difference in stellar mass between adjacent time bins and dividing by the bin size.
    
    Furthermore, we calculated the local \sigmag as the mass-averaged radial (along LoS) velocity dispersion within a patch, which represents the turbulence of the majority of the gas mass. However, this is only the pure turbulent motion $\sigma_{\rm g,turb}$ and does not account for the unresolved motion of the gas. Thus, the speed of sound in the medium $c_{\rm s}=\sqrt{\frac{\gamma k_{\rm B}T}{m_{\rm H}}}$ was added in quadrature to the velocity dispersion, $\sigma_{\rm g} = \sqrt{\sigma_{\rm g,turb}^2 + c_{\rm s}^2}$. The sound speed is on the order of a few $\kmsec$ for molecular and neutral hydrogen, and $10-15\kmsec$ for ionised gas.

    The global average of the velocity dispersion was computed as a mass-weighted average of the local patches. As with observations, the patch size was fixed and we chose to set $d_{\rm patch} = 2\,$kpc, which represents the spatial resolutions of observed, resolved, galaxies at $z\sim1-2$ (see Table~\ref{tab:data}). Varying the patch size by 1-2 kpc does not alter the results significantly. The global SFR is simply the sum of SFR of all patches within the galactic disc.
    
    We present our global averages in the coming figures as the mass-weighted mean value of the global averages calculated for the galaxy at different times and include the variation in both \sigmag and SFR as error bars. We chose to weight \sigmag by the gas mass of the corresponding phase since the mass is directly correlated to emission lines originating from the gas and is thus a good proxy for radiation flux.\footnote{We note that there is no consensus in literature on what weights should be applied when combining patches to calculate global values (see Section~\ref{sec:observations}), but we elect to compare with flux-weighting as it is one of the most common approaches and, thus, allows for a better comparison with literature data.} However, we find no significant difference between weighting global \Halpha velocity dispersions with mass, emissivity, or SFR. At most, SFR-weighting yields $\sim 10\,\%$ higher \sigmag in the feedback runs compared to mass-weighting or applying no weights. In Section~\ref{sec:mock} we perform mock observations of certain gas phases and go more into depth about phase-specific weights.

	\section{Results}
	\label{sec:results}
    We begin our analysis by visualising the simulations in Figure~\ref{fig:visual}. Shown is the mass-weighted mean gas density and temperature 50 Myr after the initial refinement period (see Section~\ref{sec:simsuite}). All simulated galaxies feature a turbulent, irregular ISM. Cold dense star forming clumps can be found in all simulations, but are most prominent when feedback is absent. In the high gas fraction models, the clouds dominate the galaxy's morphology, in agreement with observations of high-redshift galaxies \citep[e.g.][]{Elmegreen+07, Elmegreen+09, Genzel+11, Zanella+19}. At this time, the galaxies with feedback produce stars at a rate of $\sim 25\,\Msolyr$ and $\lesssim 1\,\Msolyr$ for the \texttt{fg50\_FB} and \texttt{fg10\_FB} runs, respectively, whereas the galaxies without any feedback produce 4-5 times more stars.

    Stellar feedback disrupts the dense star forming clouds and creates regions filled with hot gas in the ISM. The effect of feedback is also evident from the thickened disc structure in the vertical projection. In \texttt{fg50\_FB}, vigorous galactic outflows are present, with cold gas entrained out to 10 kpc above the disc mid-plane, as observed in local starburst galaxies \citep[see e.g.][for a review]{Veilleux+05}. While stellar feedback impacts the gaseous morphology of the galaxy, it is not clear how this manifests itself in terms of observed gas velocity dispersions, globally and locally, across different gas tracers. Quantifying this is the focus of the sections below.

		\begin{figure*}
		\centering
		\includegraphics[width=0.7\textwidth]{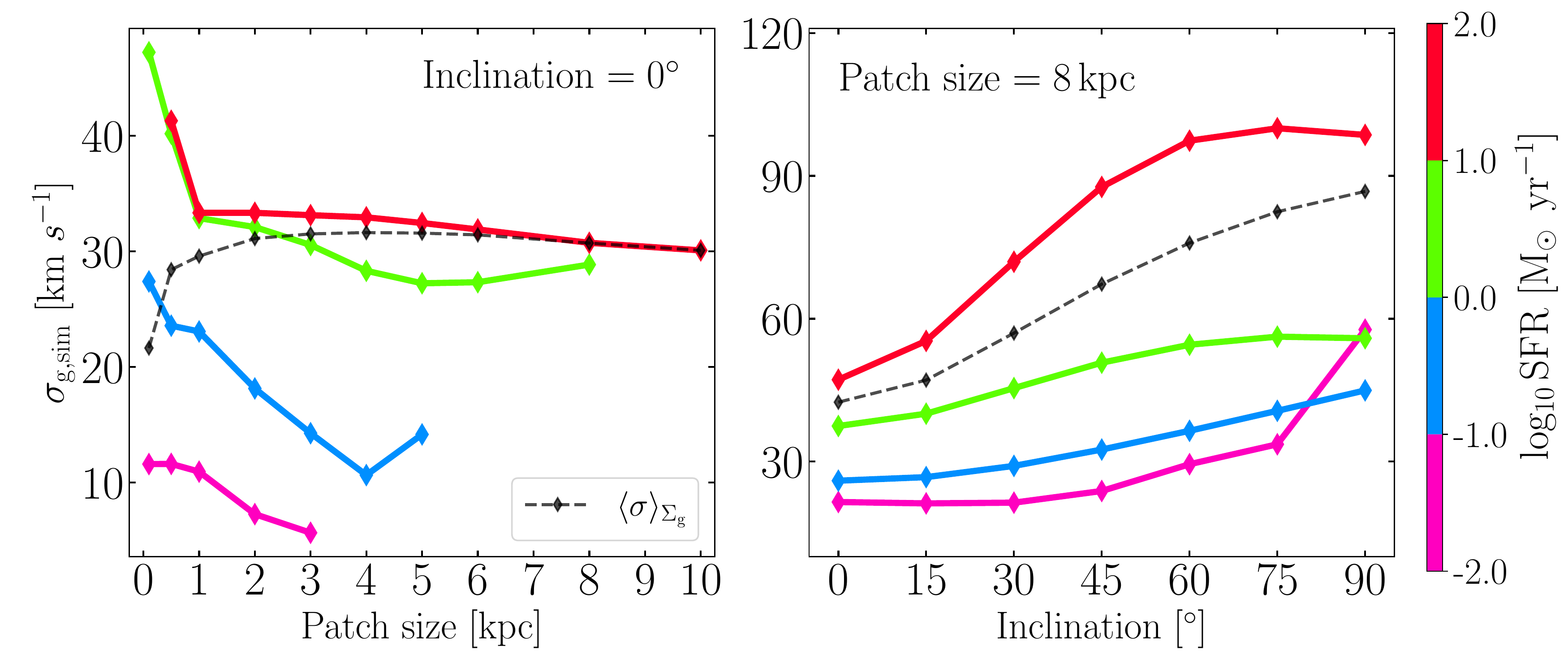}
		\caption{ The gas velocity dispersion in a range of SFR bins (according to the colorbar) as a function of the patch size (left) and the inclination angle (right). This figure shows simulated data from \texttt{fg50\_FB}, but the same relations are found for all simulation runs. The increase for smaller patch sizes in each individual SFR bin is purely due to $\Sigma_{\rm SFR}$ increasing with decreasing patch sizes. The surface density-weighted average (dashed line) shows the negligible effect of patch size on the calculated \sigmagalt. Furthermore, beam smearing severely impacts \sigmagalt, especially for large SFRs.}
		\label{fig:sigma_vs_size-ang}
	\end{figure*}
	
	\subsection{The SFR-$\sigma_{\rm g}$ relation}\label{sec:sigmasfr}
    With the morphology and behaviour of the galaxy outlined, we now focus on the \sigmasfr relation. We begin by analysing the simulated galaxies completely face-on, without any influence of observational artefacts such as beam smearing, which we investigate in Section~\ref{sec:obseff}. Figure~\ref{fig:sigmasfr} shows \sigmasfr relation for both simulated galaxies, with and without feedback, compared to the set of observational data. The local quantities (computed in patches, defined in Section~\ref{sec:reduce_data}) are represented by contours, indicating the intervals in which 30, 68, and 80\% of the data is contained. The global quantities are represented by markers with error bars covering the entire distribution of the global averages. 
	
	The plot shows an overall agreement in the \sigmasfr relation between observations and simulations for the relevant range in SFRs. A striking result it that the presence of stellar feedback does not significantly impact the simulated \sigmasfr relation, with only a broadening of the \sigmag distribution in the feedback case. The source of this broadening is the formation of the more turbulent warm ionised phase, primarily caused by supernovae, which we investigate further in Section~\ref{sec:mock}. 
	
	Furthermore, the simulations are able to reproduce two recognisable features of the observational data: the plateau at $\sigma_{\rm g}\sim 10-20\kmsec$ for SFR\,$\lesssim2\Msolyr$ and the steep increase of \sigmag$\sim 60\kmsec$ for SFR\,$\gtrsim 2 \Msolyr$. The galaxy-wide \sigmag and SFR also match well with observational data, regardless of the adopted weighting-scheme (see Section~\ref{sec:reduce_data}). However, none of our simulations are able to recover the large observed turbulent velocity in excess of $\gtrsim 60\kmsec$, as reported by many studies, including analytical work \citep[e.g.][]{KrumholzBurkhart16,KBFC18}. While this in principle could be remedied by modelling galaxies with even higher gas fractions and hence SFRs \citep[but see][]{Renaud+21}, this discrepancy can arise by not accounting for a number of key observational effects, which we explore next.

	\subsection{Observational effects}
	\label{sec:obseff}
	As discussed in Section~\ref{sec:observations}, the observational data available to probe the \sigmasfr relation is very heterogeneous and, even within the same survey, galaxies are observed at different spatial resolutions and inclinations. While inclination corrections are commonly done, even minor residual errors can impact the derived turbulent velocities (see the discussion in Section~\ref{sec:observations}). Furthermore, various gas tracers are used in the literature and it is not clear that these are directly comparable, given their vastly different origin, density etc. Our first task is thus to unpack these observational artefacts and determine their potential effect on the \sigmasfr relation.
	
	\subsubsection{Role of scale}\label{sec:scale}
	In order to quantify the effects of scale on the velocity dispersion, we analyse \sigmag as a function of patch size at fixed SFR. The relations are shown on the left panel of Figure \ref{fig:sigma_vs_size-ang} for \texttt{fg\_50FB}, which is analysed face-on. As expected, there is a clear sequence in terms of SFR and \sigmagalt, with higher SFR bins hosting higher levels of turbulence. We also notice a negative correlation between the size of the patch and \sigmagalt, for fixed values of SFR, most pronounced for SFR$\lesssim 1\,\Msolyr$. This may seem counter-intuitive given the positive correlation between velocity dispersion and size observed for molecular clouds\footnote{Note however that the Larson scaling relations are only valid on scales $\lesssim 100$ pc.} \citep[e.g.][]{Larson81,Heyer2009}. This is a direct consequence of the SFR binning, as smaller patches with the same SFR have higher surface density of star formation ($\Sigma_{\rm SFR}$), which results in higher levels of turbulence, in agreement with observations (see Appendix~\ref{app:sigma2D}). We find the same behaviour is in all simulated galaxies, regardless of the inclusion of stellar feedback or not.
	
	When computing a mass-weighted average velocity dispersion of patches as a function of scale, we recover $\sigma_{\rm g} \sim 30\kmsec$ for all scales above $\sim 1$ kpc. We note that all observations considered in this work (see Table~\ref{tab:data}) have spatial resolution $\gtrsim 500$ pc. The range of different scales present in the observations should therefore not greatly impact the inferred \sigmasfr relation, if the galaxy is observed nearly face-on.

	
	In one of the coming sections, we will show that the turbulence induced by stellar feedback is most directly detected in the ionised gas, traced by recombination lines. Feedback is believed to insert most of its turbulent energy at scales around the disc scale-height (sub-kpc scale). We analyse further the velocity dispersion in the smallest scales we probe (0.1, 0.2, 0.5\,kpc) within neutral, molecular, and ionised gas. In the presence of feedback, the neutral and molecular gas both show a noticeable increase in velocity dispersion at scales $\leq 0.5\,{\rm kpc}$ near regions with low SFR, compared to when no feedback is present. This is due to the injection of turbulent motion, by feedback, into neighbouring regions with low SFR. Furthermore, ionised gas in high SFR regions is significantly more turbulent at $\leq 0.5\,{\rm kpc}$ when feedback is present, which indicates that feedback might have a significant impact at sub-kpc scales. Our analysis reveals that the impact of feedback at various scales is complex and heavily depends on numerous factors specific to the environment observed. We leave disentangling the details of this as future work.
	
	
	
	
	

	\begin{figure*}
		\centering
		\includegraphics[width=0.8\textwidth]{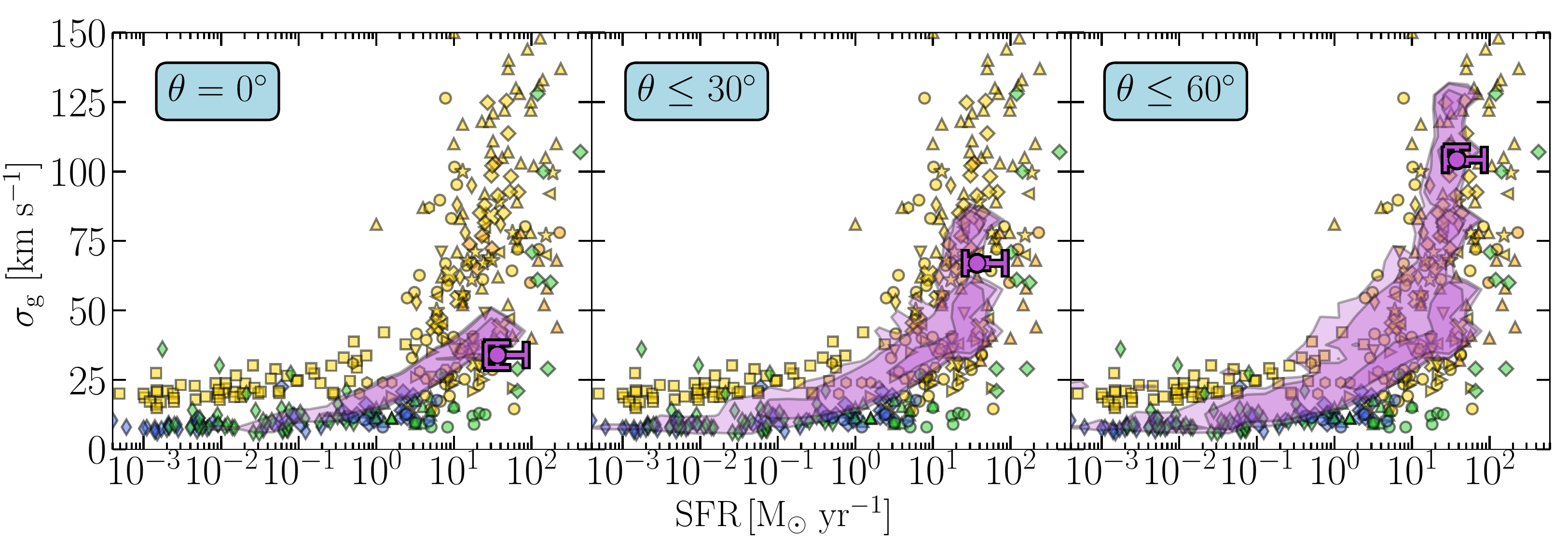}
		\caption{ The \sigmasfr relation for three particular ranges of disc inclinations, using data from the \texttt{fg\_50FB} simulation. The simulation data using a patch-approach is presented as a coloured contour and the global averages are shown as markers with error bars, as described in Section~\ref{sec:reduce_data}. The leftmost panel assumes that all galaxies are viewed head-on, while the other panels depict how the relation would look for a mixture of galaxy inclinations with $\theta\leq 30^\circ$ and $\theta\leq60^\circ$. The global averages use the inclination $\theta = 0^\circ,\,30^\circ,\,60^\circ$, respectively. Beam smearing is clearly increasing the \sigmag along LoS. Beam smearing is seen to affect patches with high SFR significantly more, but there is likely hidden variables here; the inner regions and larger patch sizes (see Section~\ref{sec:beam_smearing} for details).}
		\label{fig:sigmasfr_all_angles}
	\end{figure*}
	
		\begin{figure*}
		\centering
		\includegraphics[width=1.0\textwidth]{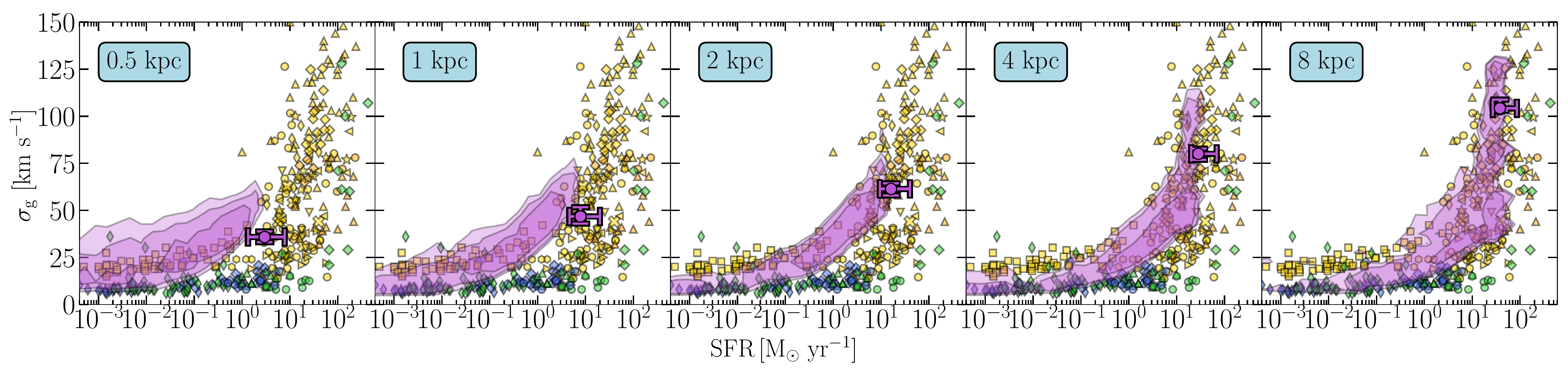}
		\caption{ The \sigmasfr relation for various patch sizes. The simulation data using a patch-approach is presented as a coloured contour and the global averages are shown as markers with error bars, as described in Section~\ref{sec:reduce_data}. The data shown here is of \texttt{fg50\_FB} with an inclination of $\theta\leq60\,^\circ$ (global averages use $\theta = 60^\circ$). The observed velocity dispersion drastically increases with larger patch sizes, which highlights the significance of beam smearing for poorly resolved observations of galaxies. As the data is highly heterogeneous, most of the high SFR data come from poorly resolved high-z galaxies. }
		\label{fig:sigmasfr_beam_smearing_patches}
	\end{figure*}

	\subsubsection{Role of beam smearing/inclination}\label{sec:beam_smearing}
	To understand the role of beam smearing (discussed in Section~\ref{sec:observations}) we measure the LoS \sigmag in patches of size 8 kpc in \texttt{fg50\_FB} for galaxy inclination angles $\theta =0-90^\circ$. Analogous to the previous section, we compute \sigmag for fixed values of SFRs as well as a mass-weighted average velocity dispersion. The results are shown in the right panel of Figure~\ref{fig:sigma_vs_size-ang}. The effect of the inclination angle is a strong function of the SFR, with \sigmag increasing by more than a factor of 2, and reaching \sigmag$>100\kmsec$ for the highest SFR bin ($\gtrsim 10\,\Msolyr$, red solid line) when the galaxy is inclined by $\theta\sim 60^\circ$. This effect is expected, as the SFR increases towards the galaxy centre where the rotation curve varies strongly with galactocentric radius. Beam smearing is here most severe for large observational patches, as velocity gradients become less resolved which broadens the velocity distribution and consequently the observed LoS \sigmagalt. 
	
	The fact that the effect of beam smearing becomes less pronounced at low SFRs has a significant impact on the resulting \sigmasfr relation, as seen in Figure~\ref{fig:sigmasfr_all_angles}. Here each panel features, from left to right, measurements for $\theta = 0^\circ,\, \leq 30^\circ,\,$ and $ \leq60^\circ$ for patch sizes $\geq 4$\,kpc. The joint effect of higher \sigmag at high inclination angles and high SFR, identified above, leads to the simulated \sigmasfr relation matching the observed trend, even up to values as large as $\sigma_{\rm g}\sim 140\kmsec$ when including galaxy inclinations up to $60^\circ$ (right panel). To demonstrate the impact of beam smearing on observed scales, we plot the \sigmasfr relation with patch sizes of $0.5,\, 1,\, 2,\, 4,\, 8$\,kpc for inclinations $\theta \leq 60^\circ$ in Figure~\ref{fig:sigmasfr_beam_smearing_patches}. This figure highlights how low observational resolution may result in severe beam smearing of the observed kinematic of gas in disc galaxies.
	
	This striking agreement suggests that insufficient inclination correction, and the accompanied beam smearing, in principle can explain the highest values of \sigmag observed for rapidly star forming galaxies. The sensitivity to even mild inclination effects (e.g $\theta \leq 30^\circ$ in Figure~\ref{fig:sigmasfr_all_angles}) illustrates the difficulty in recovering the true turbulent \sigmagalt, especially for poorly resolved high-redshift galaxies that feature highly clumpy morphologies and/or are undergoing interactions.
	
	We note that most observational data presented in this work has been corrected for inclination (see Table~\ref{tab:data}). Therefore, while some of these high \sigmag can be explained as uncertainties on the galaxy's inclination, it remains to be seen whether this explanation holds in the general case. We note that the galaxies with the highest SFRs and \sigmag are all high redshift galaxies observed in \Halphaalt. How the kinematics of this specific gas tracer differs from local tracers of molecular gas (CO) and \HIalt, and whether this can bias the \sigmasfr relation, is investigated next.

	\subsubsection{Mock observations and role of gas phase}\label{sec:mock}
	To allow for a closer comparison to observations, we next evaluate the turbulent motions for different gas tracers. Because gas phases have different characteristic densities and temperatures, their kinematics are likely to differ. This in turn complicates the simple, and common, interpretation of the observed \sigmasfr relation as a unique relation for all gas phases. For our analysis, we consider the molecular {\rm H}$_2$ gas phase (observationally traced by CO) by calculating the molecular gas fraction \citep[following][]{KMT09}, the neutral (atomic) hydrogen phase (traced by \HIalt) from the hydrogen number density, and the warm ionised phase (traced by \Halphaalt) from the \Halpha emissivity (following the process detailed in Appendix~\ref{app:Halpha}).

	\begin{figure*}
		\centering
		\includegraphics[width=0.65\textwidth]{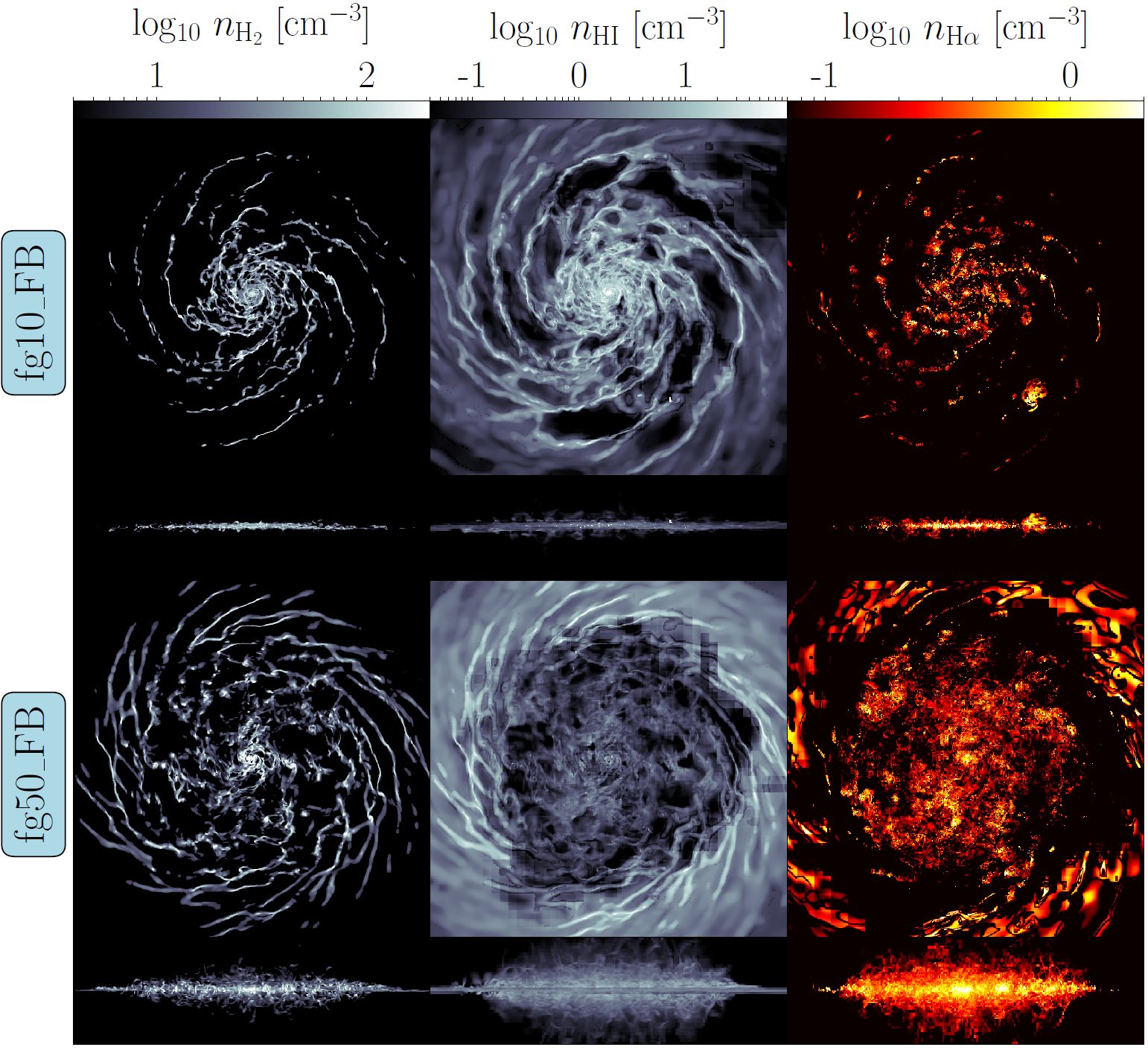}
		\caption{The mean gas number density $n$ weighted by parameters related to each phase of the different gas phases/tracers considered ({\rm H}$_2$, \HI and \Halphaalt; see Section~\ref{sec:mock}). The two simulations with feedback ($f_{\rm g} = 10\%, 50\%$) are shown 100\,Myr after feedback has been activated. The spatial size of the larger boxes is 25\,kpc (8\,kpc height for the smaller box) and each pixel of the image is 25\,pc. Notably, the molecular gas is confined within the arms and individual, massive, gas clumps. The atomic (\HIalt) gas is more evenly spread. The \Halpha tracer shows the formation of warm ionised shells, which is a consequence of stellar feedback, and highlights its impact on the ISM.} 
		\label{fig:Halpha_map_detailed}
	\end{figure*}
	
	\begin{figure*}
		\centering
		\vspace*{-3mm}
		\includegraphics[width=0.7\textwidth]{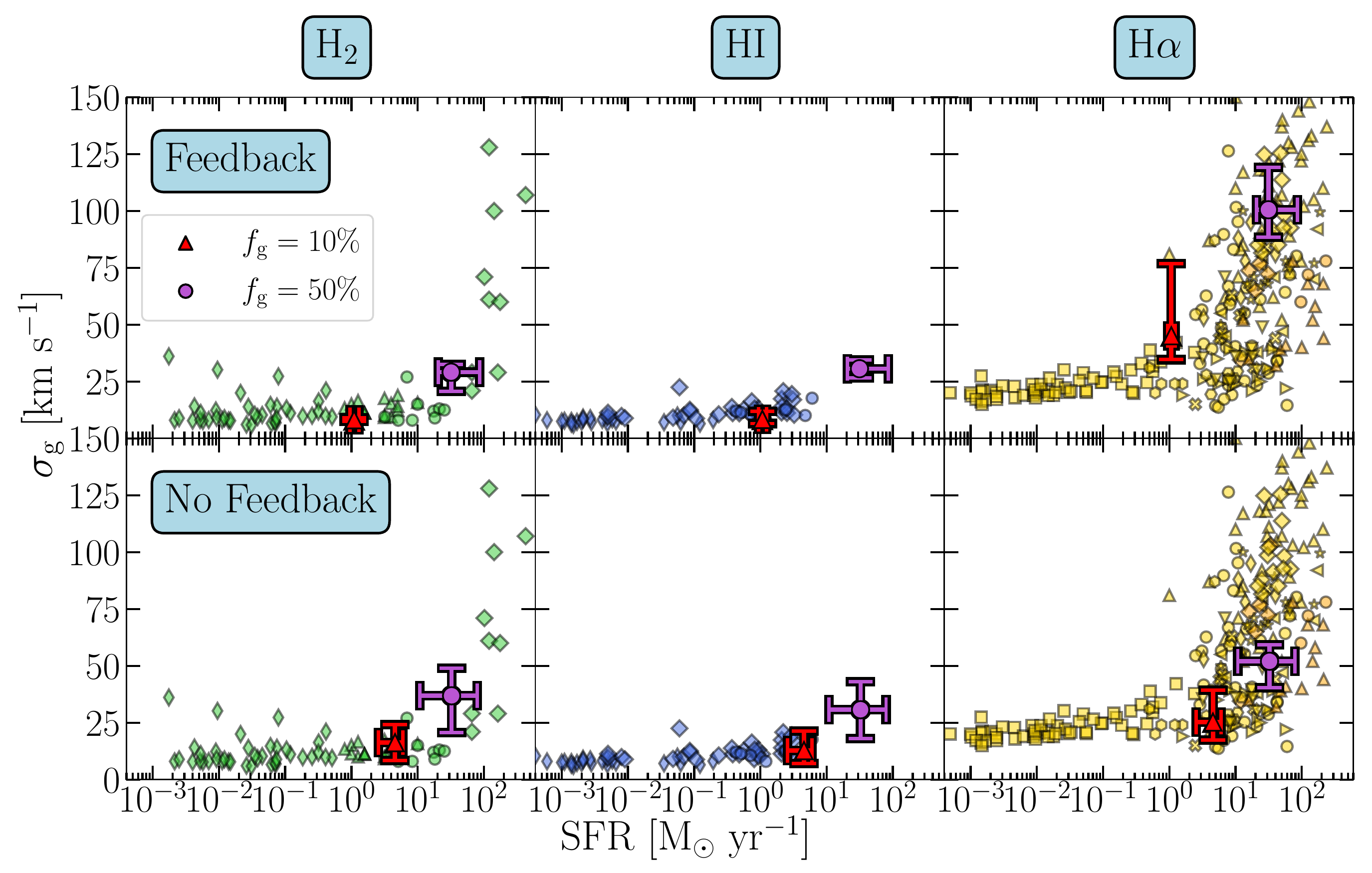} 
		\caption{ The \sigmasfr relation for the three distinct gas phase. The upper panels show the results from the simulations with feedback and the lower panels show for those without any feedback. The \sigmag of each phase is evaluated using phase-specific weights, as detailed in Section~\ref{sec:mock}. The simulations agree strikingly well with the observational data, with the exception of high SFR regions in the \HI and \Halphaalt, where there is no data due to observational limitations. Particularly, the high \sigmag observed with the \Halpha tracer is retrieved in the simulation data. }
		\label{fig:phases_mock}
	\end{figure*}

	\begin{figure*}
		\centering
		\includegraphics[width=0.7\textwidth]{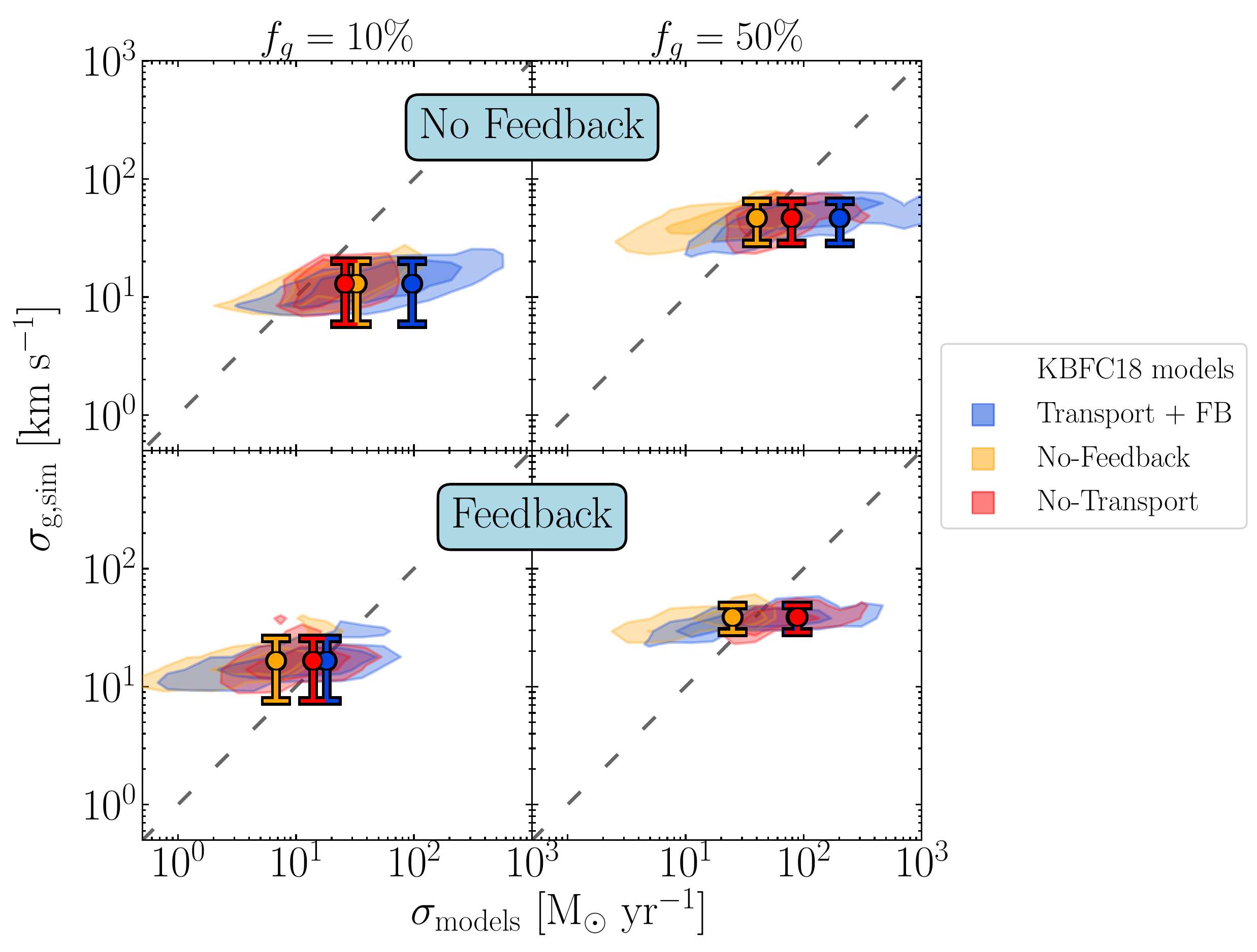}
		\caption{A direct comparison between the velocity dispersion from our simulations and the analytic models in KBFC18. The contours show the results from patches and the markers with error bars are the global averages (see Section~\ref{sec:reduce_data}). Left-side panels are the lower gas fraction galaxy and right-side show the higher gas fraction galaxy. The upper panels have been run without feedback while the lower panels are with feedback present. The dashed line shows $\sigma_{\rm g, sim}=\sigma_{\rm models}$.}
		\label{fig:comparemodels}
	\end{figure*}

A map of the gas density in each phase of the two simulations with feedback (\texttt{fg10\_FB} and \texttt{fg50\_FB}) are shown in Figure \ref{fig:Halpha_map_detailed}. The lower gas fraction galaxy is found to have a less turbulent medium, with most of the molecular and atomic gas being concentrated into its spiral arms with very little outflows seen in the edge-on view. The higher gas fraction galaxy features massive molecular gas clumps and much more intense outflows. The central 10\,kpc are dominated by molecular gas (embedded in ionised \Halpha gas) with the atomic gas phase being much more diffuse than in the low gas fraction galaxy. Both simulations feature distinct hot ionised bubbles in the \Halpha gas tracer, which are indicative of feedback driven bubbles beginning to cool down.
	
Next we evaluate the \sigmasfr relation for the three gas phases, plotted in Figure~\ref{fig:phases_mock} for all simulations. The observational data in the plots is the same as previous \sigmasfr graphs, but each panel now exclusively shows observational data for a specific gas tracer. The simulation data is here presented as weighted global averages (as explained in Section~\ref{sec:reduce_data}, using the same weights as for the maps). The figure shows that each simulated gas phase matches well with the corresponding observational data. As before, this is the case regardless of the presence of stellar feedback, but here feedback has a large impact on the velocity dispersion of the ionised gas phase, which we detail below.

Furthermore, the figure demonstrates that the turbulence is very different between different phases. The way in which the SFR and \sigmag are related is hence not (theoretically) captured by models unless they are designed for individual gas phases that can be compared to the corresponding observational data. In particular, most observational data at high SFRs and high \sigmag are high redshift galaxies in \Halphaalt. Our simulations including stellar feedback indicate that this gas phase features significantly higher levels of turbulence ($\sigma_{\rm g}>100\kmsec$) compared to the neutral ISM ($\sigma_{\rm g}<40\kmsec$) at a given SFR (see yellow symbols in Figure~\ref{fig:phases_mock}). As such, \Halpha observations are likely biased towards high \sigmag values in the \sigmasfr relation. This property was recently highlighted by \citet[][]{Girard+21}, who found that galaxies probed with \Halpha data in the DYNAMO and EDGE-CALIFA surveys had roughly 2.5 times higher \sigmag than the neutral gas, in close agreement with our findings for the simulation including stellar feedback. The EDGE-CALIFA survey of local galaxies \citep[][]{Levy+18} showed a difference between molecular and ionised velocity dispersion of a factor of roughly 2. However, some studies have suggested that galactic turbulence in ionised and molecular gas are similar \citep[e.g.][]{Ubler+18, Molina+19, Molina+20}. This discrepancy might be due to the observations tracing outflows, rather than the turbulent motions within the ISM, or beam smearing.

The \Halpha gas tracer is hence the one gas phase where we expect to find a strong signature of the effect of stellar feedback, which is required to recover $\sigma_{\rm g}\sim100\kmsec$. Furthermore, only a small fraction ($\sim 10\%$) of the available turbulent kinetic energy needs to couple to the diffuse \Halphaalt-traced gas phase in order to explain these values (see Appendix~\ref{app:halphaturb}). Without feedback in the galaxy the velocity dispersion never exceeds $\sigma_{\rm g}\sim 50\kmsec$.

\section{Discussion}\label{sec:discussion}

    \subsection{Comparison to analytic models}\label{sec:compare-analytic}
    
	We now turn to comparisons with analytic models from the literature to ascertain their potential to reproduce the observed \sigmasfr relation. Such models are useful and fast ways for studying the physics of galactic turbulence driving, such as gravitational instability and stellar feedback. Direct comparisons to full hydrodynamical simulations are beneficial for understanding any possible caveats to the simplifications made in such models. We focus on the models presented by \citet[][hereafter KBFC18]{KBFC18} \citep[see also][for earlier work]{KrumholzBurkert10,KrumholzBurkhart16} as their formalism has successfully been applied to many observational surveys of star forming galaxies \citep[including][]{Yu+19, Ubler+19, Varidel+20, Girard+21, Yu+21}. Most of these surveys use recombination lines and, thus, trace the ionised gas phase, which we find is more turbulent than the cold dense gas driving gravitational instability.

	Gravitational instabilities in KBFC18 are quantified by the $Q$ stability parameter. We will return to this topic in Section~\ref{sec:Q_stability_analysis}, but for now it suffices to note that under the approximation of an infinitesimally thin disc, this parameter for an individual galactic component (gas or stars) reads
    \begin{equation}\label{eq:Qg}
    Q_i=\frac{\kappa\sigma_i}{\pi G\Sigma_i}
    \end{equation}
    where $\kappa$ is the epicyclic frequency, $\sigma$ the radial velocity dispersion, and $\Sigma$ the surface density \citep[see][on the parameter for fluids ($i=g$), and \citealt{Toomre64} on that for stars ($i=\star$]{Safronov60, GoldreichLynden-Bell65}. Galactic discs are traditionally regarded as stable if $Q_i\geq1$.  Note, however, that this stability criterion neglects the gravitational coupling between stars and gas, the vertical structure of the disc, the effect of non-axisymmetric perturbations, and other important factors (for an overview, see sect.\ 5.2 of \citealt{RomeoFathi15} and sect.\ 4.2 of \citealt{RomeoFathi16}). The actual stability threshold is significantly higher: $Q\approx2\mbox{--}3$ \citep[e.g.][]{RomeoMogotsi17}, especially at high gas fractions (as we will show in Section~\ref{sec:Q_stability_analysis}). KBFC18 calculated a combined $Q$ parameter following $Q=f_{\rm g,Q}Q_{\rm g}$ (see their Eq. 8). We note that this couples the total $Q$ to the gas component, which we in the next section show not to be universally true, as the stellar component plays a crucial part in controlling the disc's gravitational stability.
    
    In the KBFC18 models, gas is kept in vertical pressure and energy balance by the turbulent energy and momentum injected by gravitational instability and stellar feedback, in the form of gas transport through the disc and supernovae explosions, respectively. The gas transport equations are based on the steady state disc solution by \citet[][]{KrumholzBurkert10} assuming the disc regulates itself into marginal gravitational stability.
	
	The authors derive relations for a number of interesting cases. Their fiducial model accounts for star formation, feedback and gas transport (labelled 'Transport + FB') and relates the star formation rate to the gas velocity dispersion as
	\begin{align}\label{eq:sigmasfrKB}
	    {\rm SFR} = \frac{1}{2+\beta} & \frac{\varphi_a f_{\rm sf}}{\pi G Q} f_{g, Q} v_c^2\sigma_{\rm g} \\\nonumber
	    & \times {\rm max}\bigg[\sqrt{\frac{2(1+\beta)}{3 f_{g, P}}}\varphi_{\rm mp} \frac{8\epsilon_{\rm ff}f_{g, Q}}{Q},\, \frac{t_{\rm orb}}{t_{\rm sf, max}} \bigg].
	\end{align}
	This equation involves a number of free parameters that need to be assumed or calculated. We adopt the scaling factor for turbulent dissipation rate $\eta=1.5$ and the average momentum injected by stellar feedback per unit mass $\langle p_*/m_* \rangle = 3000\kmsec$. For all of our galaxies we assume a rotation index $\beta=0$, orbital period $t_{\rm orb}=200$\,Myr and rotation curve velocity $v_c=230\kmsec$. The other model parameters are described for each simulation in Table~\ref{tab:KBFC18} along with the recommended values KBFC18 apply for these types of galaxies; local spirals and high-redshift galaxies.

	\begin{table*}
		\label{tab:KBFC18}
		\centering  
		\caption{The parameters used in the analytical models by KBFC18, shown in Eq.\ref{eq:sigmasfrKB} and \ref{eq:sigmasfrKB-FB}. The recommended values presented here for \texttt{fg10} are given in KBFC18 as 'Local spirals' and \texttt{fg50} as 'High-z'. }
		\begin{tabular}{l l l l l}
			\hline
			 & \multicolumn{2}{c}{Recommended value} &  \\
			Parameter & {\texttt{fg10}} & {\texttt{fg50}} & Description\\
			\hline 
			$f_{g,Q}$ & 0.5 & 0.7  & Fractional contribution of gas to Q   \\
			$f_{g,P}$ & 0.5 & 0.7  & Fractional contribution of gas self-gravity to mid-plane pressure  \\
			$f_{\rm sf}$ & 0.5 & 1.0  & Fractional of ISM in star-forming molecular phase  \\
            $\varphi_a$ & 1 & 3 & Offset between resolved and unresolved star formation law normalisations\\
            $\varphi_Q$ & 2 & 2 & One plus ratio of gas to stellar $Q$\\
            $\varphi_{\rm nt}$ & 1 & 1 & Fraction of velocity dispersion that is non-thermal\\
            $\varphi_{\rm mp}$ & 1.4 & 1.4 & Ratio of total pressure to turbulent pressure at mid-plane\\
            $\epsilon_{\rm ff}$ & 0.015 & 0.015 & Star formation efficiency per free-fall time\\
            $t_{\rm sf,max}$ (Gyr) & 2 & 2 & Maximum star formation time-scale\\
		\end{tabular}
	\end{table*}

	The second case we consider is that of no stellar feedback (i.e. only transport, referred to as 'No-Feedback'). This is obtained by fixing $Q$ in Eq.~\ref{eq:sigmasfrKB} to a value $Q_{\rm min}$ ($=1$) in order for the disc to remain in a marginally stable state.
	
    Finally, a feedback-only model, referred to as 'No Transport, Fixed Q', is derived by considering star formation to be the only contributing factor to the system's turbulence and has the form
	\begin{align}\label{eq:sigmasfrKB-FB}
	    {\rm SFR} & = \frac{4\eta\sqrt{\varphi_{\rm mp}\varphi_{\rm nt}^3\varphi_{\rm a}}\varphi_Q}{GQ^2 \langle p_\star/m_\star \rangle} \frac{f_{g,Q}^2}{f_{g,P}} v_c^2 \sigma_{\rm g}^2. 
	\end{align}

	We exploit the models as intended by measuring most of the parameters in the above equations directly from the simulations\footnote{Even when adopting the values suggested by KBFC18 for low- and high-z spiral galaxies (see Table~\ref{tab:KBFC18}), our conclusions remain unchanged.}. Specifically, we calculate $f_{\rm g,Q}$ (and set $f_{\rm g,P}=f_{\rm g,Q}$), $\varphi_{\rm nt}$, $\varphi_Q$ and $Q=f_{\rm g,Q}Q_{\rm g}$, using the definitions in KBFC18. As prescribed by the authors, a velocity dispersion floor of $10\,\kmsec$ is added to the 'Transport + FB' and 'No Transport, Fixed Q' models, representing the thermal broadening and turbulence contributed by feedback, which is missing in 'No-Feedback'. Ideally, analytic models should be able to self-sufficiently explain this plateau in order to understand the \sigmasfr relation, since the low-SFR regime is represented by a huge amount of galaxies. Furthermore, we have previously demonstrated (see Figure~\ref{fig:sigmasfr}) that feedback is not required to produce this plateau in the \sigmasfr relation, which indicates that this can be driven purely by gravitational instabilities.

    In Figure~\ref{fig:comparemodels} we directly compare \sigmag obtained from all four simulations and the three analytical models. As before, we show the results from local patches as well as global averages with the models (see Section~\ref{sec:reduce_data}). The dashed line indicates a one-to-one match ($\sigma_{\rm g, sim} = \sigma_{\rm model}$). For \texttt{fg10\_FB}, all analytical models fall within the globally averaged \sigmag from the simulations\footnote{By adjusting the contribution from thermal broadening, an even closer match can in principle be achieved.}. While this agreement in principle is encouraging, it limits the ability to disentangle the physics of turbulence driving (provided that the simulation results are robust). All models predict higher than measured \sigmag in \texttt{fg10\_noFB}, by as much as a factor of 4 in the most relevant model for comparison, 'No-Feedback'.

    In the $f_{\rm g}=50\%$ simulations, the 'No-Feedback' analytical model closely matches the globally averaged \sigmagalt. This is in line with the conclusions made by KBFC18 that gravity is the main source of turbulence in these systems \citep[see also][]{Yu+19}. It is noteworthy that the 'Transport + FB' model predicts values as high as $\sigma_{\rm g}>100\kmsec$, in contrast with $\sigma_{\rm g,sim}\sim 40-50\kmsec$ measured in the simulations. As discussed above (Section~\ref{sec:mock}), we argue that such high values can only be reached in tenuous gas traced by recombination lines. Related to this, less variation in \sigmag is present, for individual patches, in the simulations compared to what is predicted by the analytical models, i.e. different functional forms of the \sigmasfr relation are predicted. As a reminder, the average scaling of the observational data is $\sigma_{\rm g}\propto{\rm SFR}^{0.1}$ at low SFRs and $\sigma_{\rm g}\propto{\rm SFR}^{0.4}$ at ${\rm SFR}\gtrsim 2\Msolyr$ (see Figure~\ref{fig:data}). This is shallower than the linear relation predicted by the models including transport $\sigma_{\rm g}\propto {\rm SFR}$, but similar to the 'No-Transport' models with $\sigma_{\rm g}\propto {\rm SFR}^{0.5}$. 

    To summarise, we find that our global averages fall within what is predicted by several of the models, but this makes it difficult to resolve what physical process is driving the \sigmasfr relation. The \sigmasfr scaling is noticeably different between model and simulation, which results in models over- and under-predicting in certain patches. We note that while the above disagreements could be interpreted as a problem of the analytical models \citep[e.g.][included disc flaring in their feedback models, which resulted in a longer dissipation time of feedback turbulence and a better match to observations]{Bacchini+20}, they can also signal a problem of the simulation (e.g. missing feedback physics). Finally, it is encouraging that both simulations and analytical models point to the dominating role of gravitational instabilities in driving ISM turbulence on large scales \citep[see also][]{Agertz+09b}, but care needs to be taken when interpreting turbulence properties obtained from different gas phases.

	\begin{figure*}
		\centering
		\includegraphics[width=0.8\textwidth]{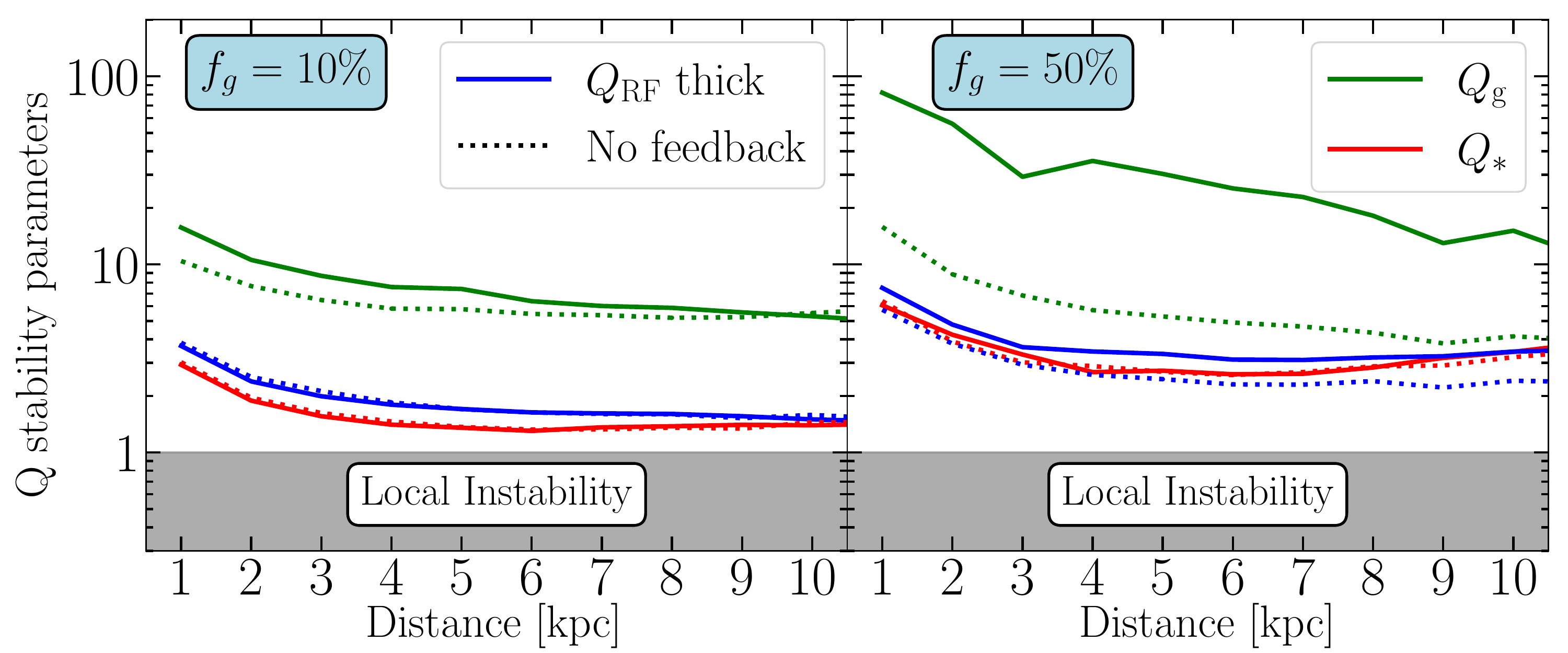}
		\caption{ Different $Q$ stability parameters plotted against the radial distance from the centre, including all simulation runs; \texttt{fg10} on the left and \texttt{fg50} on the right. The dotted lines are the runs without any feedback. The $Q$-values shown are for the gas, stars and the multi-component $Q_{\rm RF}$, explained in Section \ref{sec:Q_stability_analysis}. The values presented here were evaluated at the kpc-scale. The stars are seen to drive the instability and $Q_{\rm RF}$ shows that the lower gas fraction is quasi-stable while the higher gas fraction galaxy is marginally stable, independent of feedback heating.} 
		\label{fig:Qs}
	\end{figure*}

	\subsection{Predicting \sigmag from disc stability arguments}\label{sec:Q_stability_analysis}
	In previous sections we have found that the level of ISM turbulence present in our simulations is insensitive to the presence of stellar feedback, implying that gravitational instabilities is the main driver, at least on large (kpc) scales. This motivates a detailed stability analysis of the simulated galaxies. 

	To account for the different contributions of stars and gas on the net stability regime of a galactic disc, we adopt the multi-component parameter $Q_{\rm RF}$ introduced by \citet{RomeoFalstad13}. The general case for any number of components is defined as
	\begin{align}\label{eq:Q_RF}
	\frac{1}{Q_{\rm RF}} = \sum_i \bigg( \frac{W_i}{T_i Q_i} \bigg)
	\end{align}
	The term $T_i$ quantifies the stabilization effect of each component due to the thickness of the disc and reads
	\begin{align}\label{eq:Ts}
	T_i =
	\begin{cases} 
	1 + 0.6\hspace{1mm}\Big(\frac{\sigma_z}{\sigma_R} \Big)_i^2 \hspace{5mm}\mathrm{for} \hspace{2mm} \Big(\frac{\sigma_z}{\sigma_R}\Big)_i \leq 0.5,  \\
	0.8 + 0.7\hspace{1mm}\Big(\frac{\sigma_z}{\sigma_R} \Big)_i \hspace{5mm}\mathrm{for} \hspace{2mm} \Big(\frac{\sigma_z}{\sigma_R}\Big)_i >0.5 .  \\
	\end{cases}
	\end{align}
	Finally, the term $W_i$ is used to attribute different weights to the components:
	\begin{equation}
	W_i = \frac{2\sigma_m\sigma_i}{\sigma_m^2 + \sigma_i^2},
	\end{equation}
	here $m$ denotes the least stable component,
	\begin{align}
	    T_m Q_m = \underset{i}{\rm min}\{T_i Q_i\}.
	\end{align}
	This relation is important for differentiating between star- and gas-driven turbulence regimes, which we elaborate on in Appendix~\ref{sec:instability_driver} and show that stars are the main driver in our galaxies when $f_{\rm g} \lesssim 45\,\%$.

	The stability of galaxies can be evaluated from the $Q$ stability parameters (see Eq. \ref{eq:Qg} and \ref{eq:Q_RF}), which are shown for our simulations in Figure~\ref{fig:Qs}, as a function of the distance from the galactic centre. We present $Q$ for stars, gas and a combination of these into a multi-component $Q_{\rm RF}$ (see Section~\ref{sec:Q_stability_analysis}). Note that \Qgtext is for the total gas and that it is the cold and dense gas that contribute most to its instability; thus this analysis is not comparable with the more turbulent warm and ionised gas phase.
	
	Figure~\ref{fig:Qs} shows that the \texttt{fg10} discs are marginally stable, with $Q_{\rm RF}\sim Q_\star\sim1.5-3$ while \texttt{fg50} galaxies are slightly more stable with $Q_{\rm RF}\sim Q_\star\sim2-4$, with little dependence on feedback. The gaseous component is much higher, with $Q_{\rm g} \sim 6-7$, for \texttt{fg10} and $Q_{\rm g} \sim 10-100$ for \texttt{fg50\_FB}. However, as the stars are driving the instability in all of our galaxies, the high \Qgtext does not affect the overall stability. We find that feedback does not significantly alter the overall stability of the disc on the scales we probe it ($\sim 1\,$kpc; \citealt{Renaud+21} show that discs follow the Toomre regime of instabilities at scales $\gtrsim$ a few 100 pc). This indicates that turbulence at these scales might be driven by gravitational instability. 
	
	As shown above, we find that the discs are in a marginally unstable state ($Q_{\rm RF}\sim1.5-4$) in all simulations. As such, the stability parameter can serve as a predictor (within a factor of 2) for the gas velocity dispersion. Re-writing the $Q_{\rm RF}$ equation to solve for the gas velocity dispersion
	\begin{align}\label{eq:sigmaRF}
	\sigma_{\rm RF} &= \sigma_\star \sqrt{2\,\frac{\Sigma_{\rm g}}{\Sigma_\star} \frac{T_\star}{T_{\rm g}} \big( T_\star Q_\star / Q_{\rm RF} -1 \big)^{-1} - 1},
	\end{align} 
	where we have assumed stars drive the instability. This predictive equation is easily applicable to data of disc galaxies where the velocity dispersions and surface densities of the gas and stars are known. In particular, for our analysis we measured $\Sigma_{\rm g},\,\Sigma_*,\,\sigma_{\rm g},\,\sigma_*,\,Q_*$ from the simulations and then assumed $T_\star\approx 1.2,\ T_{\rm g}\approx 1.5$ and $Q_{\rm RF}\approx2.2$ \citep[as measured for local disc galaxies][]{Leroy+08, RomeoFalstad13, RomeoMogotsi17}.
	
\begin{figure}
		\centering
		\includegraphics[width=0.45\textwidth]{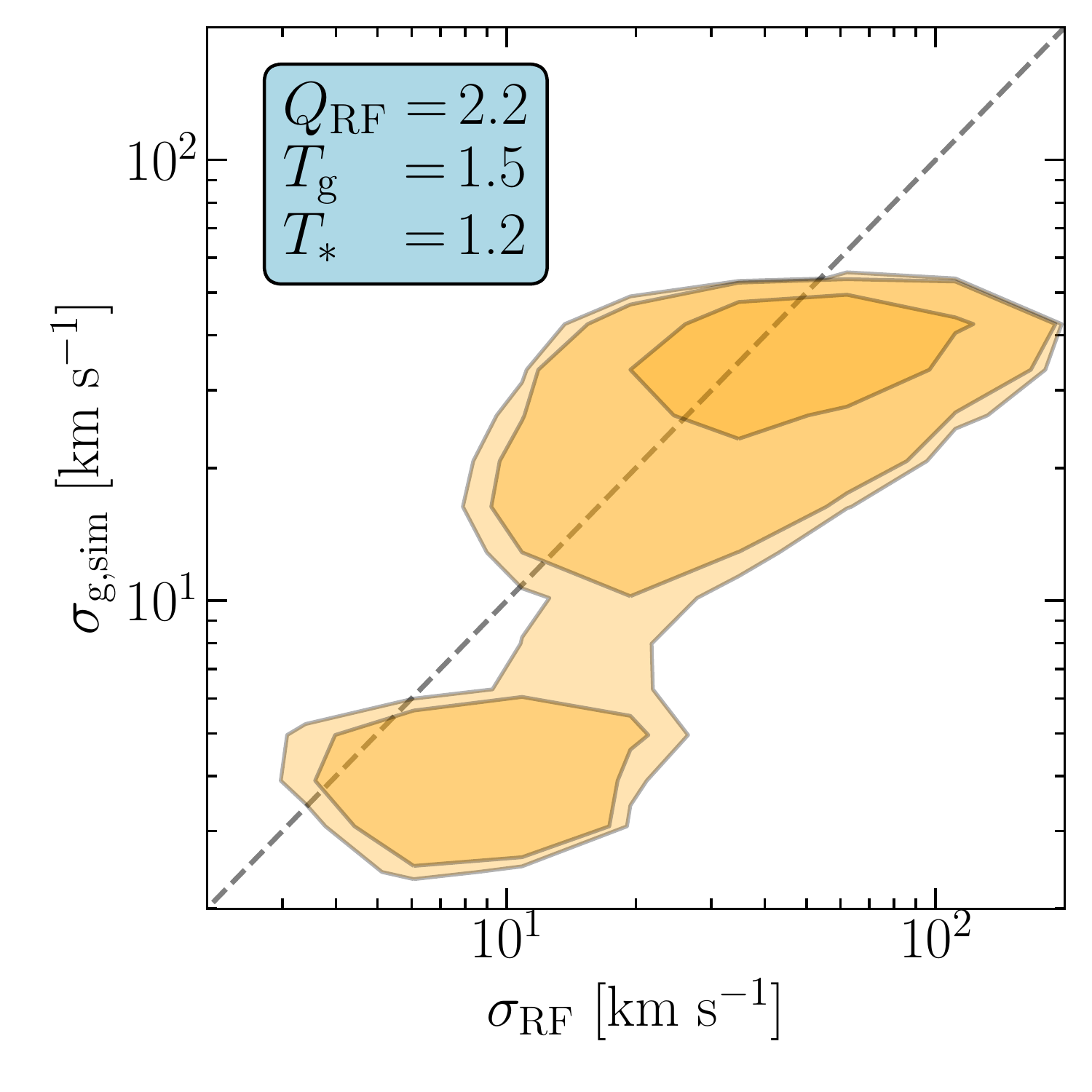}
		\caption{A direct comparison between the \sigmag calculated from simulations and the \sigmag predicted by a combined $Q$ parameter recipe from \citet{RomeoFalstad13}, see Eq. \ref{eq:sigmaRF}. We find an encouraging match between $\sigma_{\rm RF}$ and the simulation data. The $\sigma_{\rm RF}$ values are seen to cluster around $50\kmsec$ and $10\kmsec$. The lower values arise due to the significantly different stability conditions occurring in the outer region of the galaxy, where other physical processes might balance against gravitational collapse. }
		\label{fig:sigma_from_Q}
	\end{figure}

	In Figure~\ref{fig:sigma_from_Q} we compare this analytic equation with our simulated \sigmag for a marginally unstable disc. This figure demonstrates that the $Q_{\rm RF}$ approximation can predict the level of turbulence in disc galaxies. Furthermore, we discern two regions for the contoured data; below and above $\sigma_{\rm g, sim} \sim 6-10\kmsec$. The lower \sigmag mainly represent the outskirts of the galaxy, $\gtrsim10\,$kpc. This might indicate that different processes might be driving the turbulence in the outer and inner parts of the galaxy. 
	
	\section{Conclusions}\label{sec:conclusions}
	
	In this paper we present hydrodynamical simulations of isolated disc galaxies in order to evaluate how observational parameters shape the \sigmasfr relation. The available data from observations is highly heterogeneous and the various effects need to be carefully considered in order to analyse the underlying driver of this relation. Then we compare the velocity dispersion predicted by theoretical models with our simulations. Our findings can be summarised as follows:
	\begin{enumerate}
		\item Our simulations reproduce key features in the observed \sigmasfr relation: the plateau at $\sigma_{\rm g}\sim 10\kmsec$ for SFR$\lesssim 2\Msolyr$ and the exponential growth towards larger $\sigma_{\rm g}$ for SFR$\gtrsim 2\Msolyr$. This result is independent of whether stellar feedback is included or not, and is hence an outcome of how galactic discs regulate their gravitational stability (see point (v) below).
		\item Our most turbulent galaxies, \texttt{fg50}, reach at most $\sigma_{\rm g}\sim 50 \kmsec$ at ${\rm SFR}\sim10-50\,\Msolyr$, while the velocity dispersions of high-redshift galaxies have been observed to reach $\sigma_{\rm g}\sim 100 \kmsec$. We demonstrated that $\sigma_{\rm g}>50\kmsec$ can partially be explained by beam smearing. If unaccounted for, or accounted for insufficiently, even galaxy inclinations as low as $\theta=30^\circ$ can increase the observed velocity dispersions by factors of several. High-redshift galaxies are poorly resolved, and tend to have complex morphologies due to fragmentation and merging. It is therefore possible that beam smearing leads to over-estimated levels of gas turbulence in those galaxies.
		\item While stellar feedback does not change the \sigmasfr relation for the total gas content, it affects the level of turbulence in the warm ionised gas phase, here traced by the \Halpha transition. In this gas phase, feedback significantly increases $\sigma_{\rm g} \sim 50\ \kmsec$ to $\sigma_{\rm g} \sim 100\ \kmsec$. As stated above, this is in good agreement with what is found in high-redshift galaxies, especially given that those observations have predominantly surveyed them in \Halphaalt. However, it further highlights that different gas tracers do not trace the same gas kinematic and, thus, that the available \sigmasfr data does not follow one universal relation. Furthermore, we formulated a simple equation of the turbulent energy budget (see Appendix~\ref{app:Halpha}) and find that only 10\% of the total turbulent energy needs to be in the warm ionised gas phase (traced by \Halphaalt) in order to reproduce the high $\sigma_{\rm H\alpha}$ reported. 
		\item Analytic models for the \sigmasfr relation for gravitational instability and stellar feedback, taken from literature, can in principle reproduce the observed high velocity dispersion in rapidly star forming high-redshift galaxies. However, the parameters adopted in these models can be uncertain, and the functional form of the predicted \sigmasfr relation does not match that of the simulations.
		\item The simulated galaxies are, when both stars and gas are accounted for, naturally drawn towards a state of marginal stability, with a Q stability level of $Q\sim 1 - 3$, \emph{independent} of the source of the heating (galactic dynamics or stellar feedback). As mentioned above, this appears to be a fundamental property of galactic discs. As such, a multi-component $Q$ parameter is a valuable analysis tool for predicting levels of gas turbulence in disc galaxies. We exploited the combined $Q_{\rm RF}$ \citep[][]{RomeoFalstad13} to solve for \sigmag and found that even if assuming certain parameter values, this analytic relation can reproduce the gas turbulence within our simulations. Thus, it could possibly be applied to observational data to predict the turbulence of isolated disc galaxies.
	\end{enumerate}

In this work we have studied isolated disc galaxies, which means that cosmological effects, such as accretion and mergers, are not present. These effects would mix the ISM and possibly give rise to more turbulent motion. Indirectly, accretion would help sustain the turbulence in the galaxy over a longer period of time by continuously supplying gas for consistent star formation (and stellar feedback). Furthermore, a number of feedback process present in galaxies are not accounted for (e.g. feedback from active galactic nuclei, cosmic rays). Despite neglecting these factors, we are able to match the observed \sigmasfr relation, which implies that disc galaxies do not necessarily require these effects to remain in a marginally stable state or sustain their high levels of turbulence. 
	
	\section*{Acknowledgements}
    We would like to acknowledge that these simulations were made possible using computational resources at LUNARC, the centre for scientific and technical computing at Lund University, on the Swedish National Infrastructure for Computing (SNIC) allocation 2018/3-314, as well as allocation LU 2018/2-28. Storage resources part of allocation SNIC 2020/6-22 were used to store the data for longer term use. OA and FR acknowledge support from the Knut and Alice Wallenberg Foundation. OA acknowledges support from the Swedish Research Council (grants 2014-5791 and 2019-04659). We are grateful for the useful comments from Filippo Fraternali, Mahsa Kohandel, Cecilia Bacchini, and Juan Molina.  We thank the referee for their insightful comments.
    
	For this project, we have made use of numpy \citep{Harris+20}, matplotlib for {\small PYTHON} \citep{Hunter+07}. Visualisation of the simulation volume and handling the data was done using the {\small YT} project \citep{Turk+11}.

	\section*{Data availability}
	The data underlying this article will be shared on reasonable request to the corresponding author.
	
	
	
	
	\bibliographystyle{mnras}
	\bibliography{references.bbl}
	
	
	
	\newpage
	\appendix
	
	\section{Observational data}
	This Appendix contains a Table, Table\,\ref{tab:data}, with a compilation of literature references, from which we collected the values of \sigmag and the SFR. Each reference is associated with a marker, which are used in figures throughout this article. Furthermore, we have summarized some of the more useful properties of these observations in the Table: the survey/telescope employed, the tracer used to calculate \sigmagalt, the spatial area probed, the approximate redshift (local or high-redshift) and whether the observations were corrected for the inclination of the galaxy.
	
	\begin{table*}
		\caption{The complete literature library of the observational data of SFR and \sigmagalt. The data compiled by us is from the SFR derived by \citet{Walter+08} and the (instrument free) \sigmag is from \citet{Tamburro+09}. This table relates each reference to a marker, which are used throughout this paper when plotting quantities from that reference. Note, the observational spatial resolution and redshifts are approximate and the typical medians of the respective sample. Galaxies noted with a redshift 'local' were observed at redshifts much smaller than 1, but the precise redshift is commonly not specified. The data of this table is very heterogeneous, i.e. the observations are of different types of disc galaxies (e.g. normal, dwarf, merged galaxies) with different crucial properties (e.g. mass, morphology). A more thorough explanation of the data and its heterogeneity can be found in Section~\ref{sec:observations}.} 
		\centering  
		{\footnotesize 
			\begin{tabular}{p{40mm}|p{20mm}|p{15mm}|p{15mm}|l|p{16mm}|r}
				\hline\hline
				Reference name & Survey/Instrument & Tracer & Spatial resolution & Redshift & Beam corrected & Mark \\
				\hline 
				\vspace{-2.5mm}\citet{Alcorn+18} & ZFIRE	&	H$\alpha$	&	$\sim4$\,kpc	&	2.2	& Yes &  \includegraphics[width=0.025\textwidth]{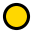} \\\hline
				\vspace{-2.5mm}\citet{Cresci+09} & SINFONI/ SINS	&	H$\alpha$	&	$\sim3$\,kpc	&	2	& Yes & \includegraphics[width=0.025\textwidth]{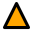} \\\hline
				\vspace{-2.5mm}\citet{DiTeodoro+16} & KMOS &	H$\alpha$	&	$\sim5$\,kpc	&	1	& Yes & \includegraphics[width=0.025\textwidth]{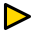}\\\hline
				\vspace{-2.5mm}Compiled by us (see Table description) & THINGS &	HI	&	 $0.1-0.5\,$kpc	&	Local	& Yes & \includegraphics[width=0.025\textwidth]{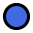}\\\hline
				\vspace{-2.5mm}\citet{Epinat+09} & SINFONI	&	H$\alpha$	&	$\sim5$\,kpc	&	$1.2-1.6$	& Yes & \includegraphics[width=0.025\textwidth]{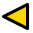} \\\hline
				\vspace{-2.5mm}\citet{Genzel+11} & SINS	&	H$\alpha$	&	$\sim2$\,kpc	&	2.3	& Yes & \includegraphics[width=0.025\textwidth]{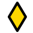} \\\hline
				\vspace{-2.5mm}\citet{Girard+21} (ionised gas) & SINS	&	H$\alpha$	&	$\sim1$\,kpc	&	  Local (\& 1-2)	& Yes  & \includegraphics[width=0.025\textwidth]{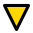} \\\hline
				\vspace{-2.5mm}\citet{Girard+21} (molecular gas) & SINS	&	CO	&	2\,kpc	&	Local (\& 1-2)	& Yes  & \includegraphics[width=0.025\textwidth]{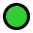} \\\hline
				\vspace{-2.5mm}\citet{Ianjamasimanana+12} & THINGS	&	HI	&	sub-kpc	&	Local	& Yes & \includegraphics[width=0.025\textwidth]{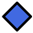} \\\hline
				\vspace{-2.5mm} Compiled by \citet{KBFC18} & SIMBAD	&	CO/ HCN	&  sub-kpc	&	Local	& No & \includegraphics[width=0.025\textwidth]{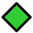} \\\hline
				\vspace{-2.5mm}\citet{Law+09} & Keck/ OSIRIS	&	H$\alpha$, [OIII]	&	1.2\,kpc	&	$\sim2.3$	& No & \includegraphics[width=0.025\textwidth]{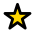} \\\hline
				\vspace{-2.5mm}\citet{Lehnert+13} & SINFONI	&	H$\alpha$, [NII]	&	$\sim5$\,kpc	&	$1-3$	& No & \includegraphics[width=0.025\textwidth]{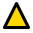} \\\hline
				\vspace{-2.5mm}\citet{Lemoine-Busserolle+10} & VLT/  SINFONI	&	UV (H$\beta$ / [OIII])	&$\sim4$\,kpc	&	$3.3$	& Yes & \includegraphics[width=0.025\textwidth]{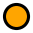} \\\hline
				\vspace{-2.5mm}\citet{Levy+18} & EDGE-CALIFA	&	CO	&	sub-kpc	&	Local	& Yes & \includegraphics[width=0.025\textwidth]{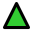} \\\hline
				\vspace{-2.5mm}\citet{Moiseev+15} & 6-m tele. SAO RAS	&	H$\alpha$	&	sub-kpc	&	Local	& No & \includegraphics[width=0.025\textwidth]{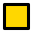} \\\hline
				\vspace{-2.5mm}\citet{Nguyen-Luong+16} & CfA 1.2\,m telescopes	&	CO	&	sub-kpc	&	Milky Way	& No & \includegraphics[width=0.025\textwidth]{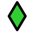} \\\hline
				\vspace{-2.5mm}\citet{Patricio+18} & MUSE	&	[OII]	&	sub-kpc	&	$0.6-1.5$	& Yes & \includegraphics[width=0.025\textwidth]{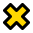} \\\hline
				\vspace{-2.5mm}\citet{Stilp+13} & THINGS\,\,and VLA-ANGST	&	HI	&	0.2\,kpc	&	Local	& Yes & \includegraphics[width=0.025\textwidth]{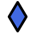} \\\hline
				\vspace{-2.5mm}\citet{Varidel+16} & SDSS	&	H$\alpha$	&	$\sim1$\,kpc	&	Local	& Yes & \includegraphics[width=0.025\textwidth]{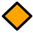} \\\hline
				\vspace{-2.5mm}\citet{Wisnioski+11} & WiggleZ	&	UV, [OII]	&	$0.8$\,kpc	&	$1.3$	& No & \includegraphics[width=0.025\textwidth]{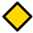} \\\hline
				\vspace{-2.5mm}\citet{Yu+19} & MaNGA	&	\Halpha	&	1\,kpc	&	0.01 - 0.15	& Yes & \includegraphics[width=0.025\textwidth]{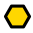} \\\hline
				\vspace{-2.5mm}\citet{Zhou+17} & SAMI	&	\Halpha	&	2.5\,kpc	&	0.05	& Yes & \includegraphics[width=0.025\textwidth]{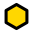} \\
				\hline
			\end{tabular}
		\label{tab:data}
		}
	\end{table*}
	
	\section{Star formation density} \label{app:sigma2D}
\begin{figure}
		\centering
		\includegraphics[width=0.45\textwidth]{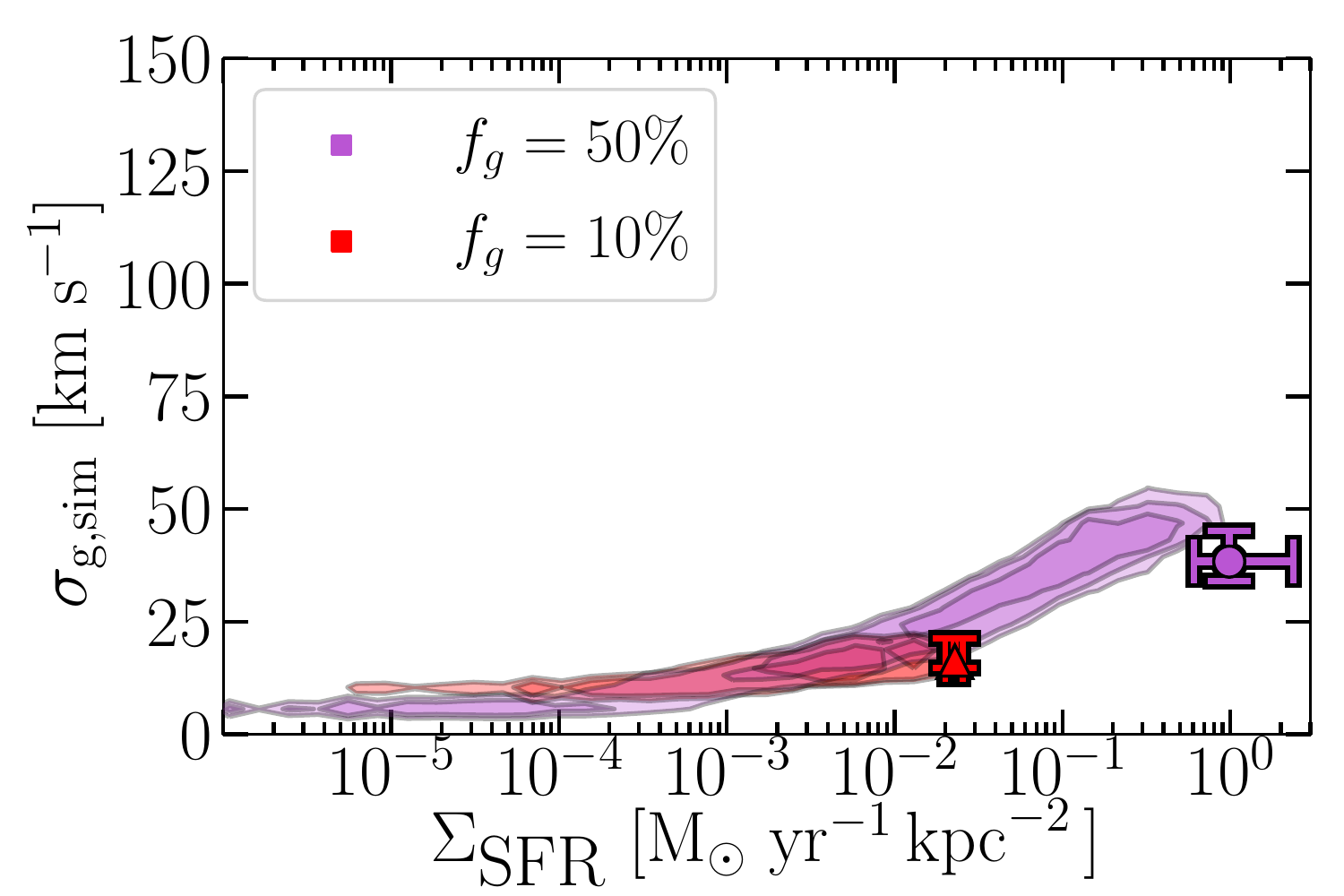}
		\caption{ The $\sigma_{\rm g}-\Sigma_{\rm SFR}$ relation for the two feedback simulations, using data with patch sizes $\leq 8$\,kpc and no inclination. The contours show the patch-based data and the global values are represented by markers with error bars, as described in Section~\ref{sec:reduce_data}. There is a clear correlation between the turbulence and star formation density, which exhibits similar properties as the \sigmasfr relation. }
		\label{fig:sigma-2Dsfr}
	\end{figure}

	In order to get a view of how star formation affects turbulence regardless of patch size, we plot \sigmag against $\Sigma_{\rm SFR}$ in Figure~\ref{fig:sigma-2Dsfr}. This relation is in agreement with \citet[][]{Orr+20} and is essentially the same as Figure~\ref{fig:sigmasfr}, but takes into account the patch size of each observed point and is, by comparison, less dispersed. As discussed in Section~\ref{sec:scale}, two patches of different sizes with identical SFR will exhibit different densities of star formation. Thus, the smaller patch will have a higher column star formation density, which leads to more intense turbulence since star formation events increase \sigmagalt.

	\section{Analytic recipe for \Halpha\ emissivity}\label{app:Halpha}
	We evaluated the warm ionised phase using emissivity equations from theoretical work, assuming the phase is well-traced by the \Halpha transition. We only consider the \Halpha emissivity contribution from the $3p \rightarrow 2s$ transition in hydrogen, since the contribution to the emission intensity of other transitions is comparably small. Emissivity has the general shape
	\begin{equation}
	\epsilon_{\rm H\alpha} = n_{\rm e} n_{\rm H} h \nu q,
	\end{equation}
	where $\nu$ is the light frequency, $h$ is the Planck constant and $q$ is the emission rate. From \ramses we can extract the electron density $n_{\rm e}$ and hydrogen density $n_{\rm H}$ of the gas. The energy of the \Halpha emission is $h\nu= 3.026 \times 10^{-12}\ {\rm ergs}$. For this analysis we consider both the emissivity coming from collisional excitation and recombination. The collisional excitation rate is given by
	\begin{equation}
	q_{\rm coll}(T) = \frac{1.3 \times 10^{-6} }{T^{0.5}} \Bigg( \frac{T}{11.2} \Bigg)^{0.305} \times \exp{\Bigg( \frac{-h\nu}{kT} \Bigg)}
	\end{equation}
	and was determined by fitting data from \citet{Callaway+87}. The recombination rate \citep[following the formalism in][]{Dijkstra17} is given by
	\begin{equation}
	q_{\rm recom} = \epsilon^B_ {\rm H\alpha}(T) \alpha_{\rm B}(T),
	\end{equation}
	where 
	\begin{align}\nonumber
	\epsilon^B_ {\rm H\alpha}(T) =\ & 8.176\times 10^{-8}- 7.46\times10^{-3}\, \log_{10}(T/10^4)\\
	& + 0.45101\, (T/10^4)^{-0.1013}
	\end{align}
	is the emitted rate of \Halphaalt; the shape is fitted using observational values from \citet[][]{StoreyHummer95}. Finally, 
	\begin{equation}
	\alpha_{\rm B}(T) = 2.753\times10^{-14} \, \bigg(\frac{315614}{T}\bigg)^{1.5} \, \Bigg(1 + \bigg( \frac{315614}{T}\bigg)^{0.407}\Bigg)^{-2.42}
	\end{equation}
	is the probability that the cascade from hydrogen recombination, called case B recombinations, emits an \Halpha photon \citep[fit from][]{HuiGnedin97}. The two $q$ values are then added linearly to calculate the total emissivity.

	\section{\Halpha turbulent energy input}
	\label{app:halphaturb}
	We showed in Section~\ref{sec:mock} that it is only through the warm ionised phase that we can achieve the large \sigmag observed at high-z, we next quantify the required energy input in order to achieve this high turbulence in the \Halpha tracer. By applying the simplistic assumption that turbulent energy in the ISM is conserved between all phases, we can derive the velocity dispersion of the ionised phase. Essentially, we assume that a fraction of the total turbulent energy is in the warm ionised phase, $F_{\rm H\alpha}$. This yields
	\begin{equation}\label{eq:sigma-Halpha}
	\sigma_{\rm H\alpha} = \sigma_{\rm g} \sqrt{ \frac{F_{\rm H\alpha}}{f_{\rm m,\, H\alpha}} },
	\end{equation}
	where $f_{\rm m, H\alpha}$ is the fraction of mass detectable in \Halphaalt. Particularly, $F_{\rm H\alpha}$ is a free parameter which tells us how much energy needs to be injected into the ionised gas phase in order to reach the $\sigma_{\rm H\alpha}$ observed.
	
\begin{figure}
		\centering
		\includegraphics[width=0.45\textwidth]{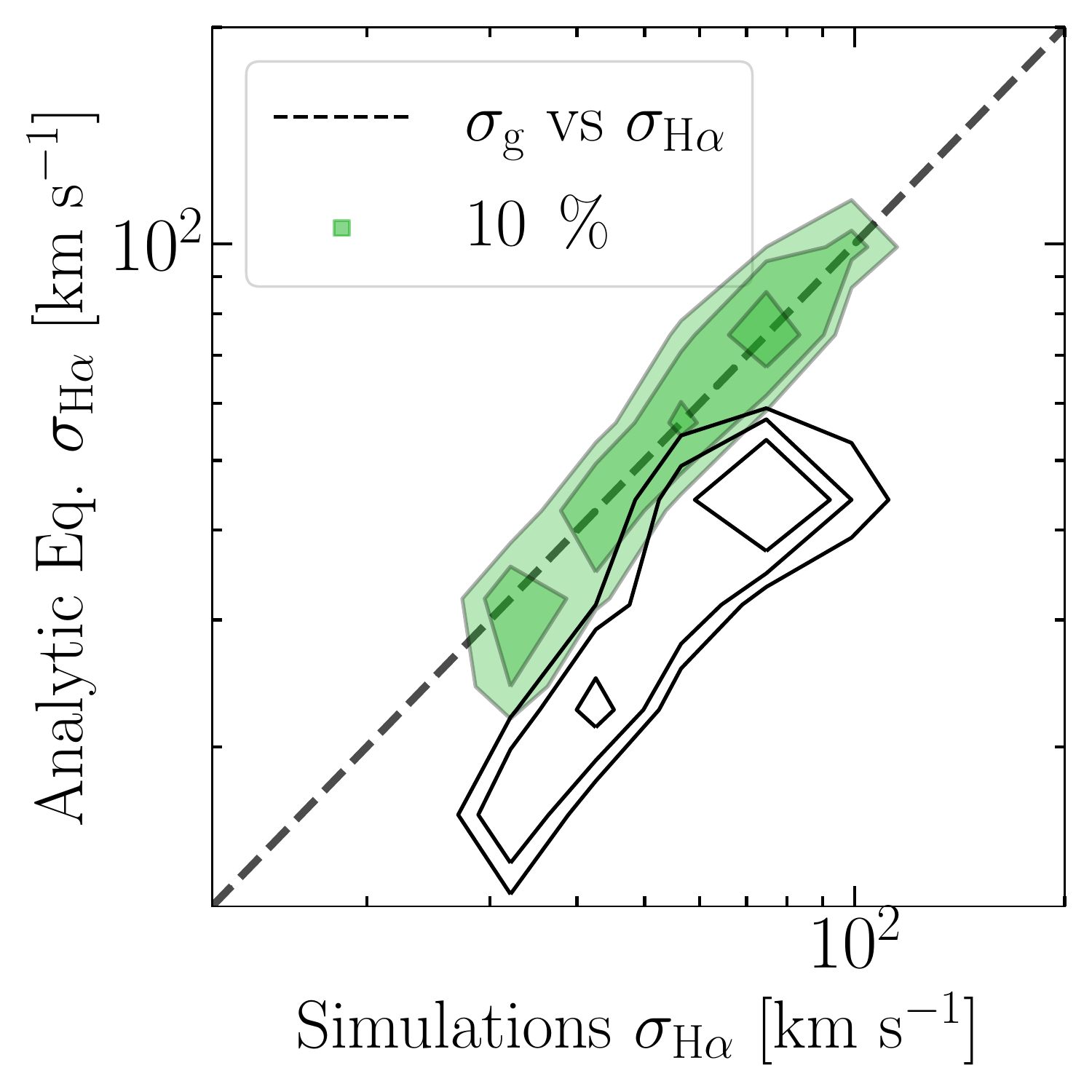}
		\caption{ The relation between the $\sigma_{\rm H\alpha}$ from a simplistic relation, shown in Eq.~\ref{eq:sigma-Halpha}, and $\sigma_{\rm H\alpha}$ calculated from simulations. This plot uses data from \texttt{fg\_50FB} and defines the \Halpha phase to temperature cuts ($6\times10^3\,{\rm K}\leq T \leq 2\times10^4$\,K), in order to calculate the gas fraction of the phase. The dotted diagonal line shows where the equation and the simulation agrees. Different factors for the energy contributed by \Halpha were tried and the best agreement was found to be $F_{\rm H\alpha}\approx 10\%$. As a comparison, the black contours show the relation of the calculated \sigmag (of all the gas) and $\sigma_{\rm H\alpha}$ from the simulations.}
		\label{fig:sigma-Halpha}
	\end{figure}

	In Figure~\ref{fig:sigma-Halpha}, we plot this equation against the $\sigma_{\rm H\alpha}$ using data from the \texttt{fg\_50FB} simulation (showing only the patch-based approach). The comparison shows a remarkable agreement in values and scaling between analytic equation and simulation. The best match is given when roughly 10\% of the turbulent energy is in \Halphaalt, which is assumed a global factor here, but might vary in the local environment of the patch. Furthermore, a comparison of how \sigmag for the total gas scales with $\sigma_{\rm H\alpha}$ (calculated) is plotted as a dashed region to illustrate how the calculated \sigmag is transformed. Note that in order to calculate the mass from our simulations we have here applied temperature cuts ($6\times10^3\,{\rm K}\leq T \leq 2\times10^4$\,K) to evaluate the \Halpha tracer (instead of the emissivity analytic equations), which was tested and found to give very similar \sigmag to our more detailed approach in Section~\ref{sec:mock}.
	
	\section{Is instability driven by gas or stars?}\label{sec:instability_driver}

\begin{figure*}
		\centering
		\includegraphics[width=0.9\textwidth]{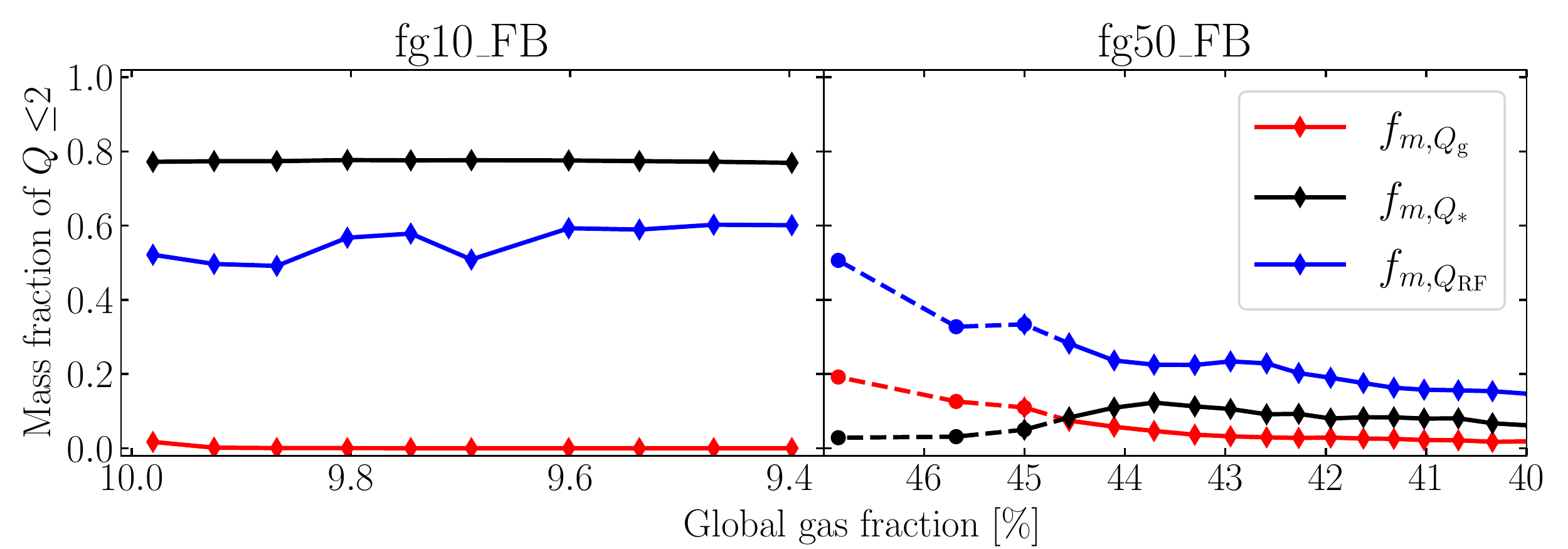}
		\caption{ The fraction of mass in a marginally stable state disc ($Q\leq 2$) as a function of the total gas fraction of the galaxy. Dashed lines (seen for gas fractions $f_{\rm g}\geq45\,\%$) correspond to the region in which the instability is dominated by the gas. A clear transition from gas- to star-driven instability is seen to occur early in the evolution of the \texttt{fg50\_FB} galaxy. For the most part, stars are the main drivers of instability in our galaxies.}
		\label{fig:fQg_fg}
	\end{figure*}

	The instability of the galactic disc is driven by its two components, gas and stars. Investigating the main driver of this instability is useful in the case that gravitational instability drives turbulent gas motion in the disc. In order to evaluate each components contribution to drive instability, we calculated the mass fraction of gas which is marginally unstable ($Q\lesssim 2$) in the disc and plot it against the gas fraction, seen in Figure~\ref{fig:fQg_fg}. We analyse the $Q$ stability parameters for the gas, stars and the combined $Q_{\rm RF}$ for a thick disc. The region in which gas dominates the instability is shown as a dashed line and is determined from Eq.~\ref{eq:Q_RF}. The dependence of the stability with gas fraction (and thus, time) is evident, as the gas mass fraction that is gravitational unstable is significantly higher in the earlier lifetime of the galaxy. Gas dominates the instability for a very limited time in the high gas fraction, ending after only a small fraction of the gas had been turned into stars. Furthermore, for the galaxies without feedback the gas dominated the disc instability far longer. This makes sense from the perspective that stellar feedback heats up the gas, while the stars remain kinematically cold \citep[see][]{vanDonkelaar+21}. Note that the decision to consider values corresponding to a marginally stable disc is motivated by Figure \ref{fig:Qs} (see also the discussion in Section~\ref{sec:compare-analytic}), in which it is evident that the higher gas fraction galaxy does not go far below this threshold. 
	
	\bsp	
	\label{lastpage}
\end{document}